\documentclass[a4paper]{article}

\usepackage{url}
\usepackage{algorithm, algorithmic}
\usepackage{amsmath,amsfonts,amssymb,latexsym,color}
\usepackage{graphicx}

\newtheorem{theorem}{Theorem}[section]
\newtheorem{proposition}[theorem]{Proposition}
\newtheorem{lemma}[theorem]{Lemma}
\newtheorem{corollary}[theorem]{Corollary}
\newtheorem{definition}[theorem]{Definition}
\newenvironment{proof}{\begin{genproof}}{\end{genproof}}
\newcommand{\qed}{}
\newenvironment{genproof}[1][]{\begin{trivlist}\item \textbf{Proof#1:} }{\nolinebreak\qquad\nolinebreak\framebox(5,5)[lb]{}\nolinebreak\end{trivlist}}

\newcommand{\etal}{et al.~}

\newcommand{\ie}{i.e.~}

\newcommand{\wloge}{w.l.o.g.~}
\newcommand{\wlogen}{w.l.o.g.}
\newcommand{\mc}{\mathcal}

\newcommand{\dist}{\mathrm{dist}}

\newcommand{\Nei}{\ensuremath{\mathcal N}}

\newcommand{\edge}[2]{\ensuremath{(#1, #2)}}

\newenvironment{titemize}{\begin{itemize}\setlength{\parskip}{-0.1cm}}{\end{itemize}}

\DeclareSymbolFont{AMSb}{U}{msb}{m}{n}
\DeclareSymbolFontAlphabet{\mathbb}{AMSb}
\newtheorem{Req}{Requirement}

\newcommand{\metricd}{metric dimension}
\newcommand{\metricdp}{\textsc{Metric Dimension}}
\newcommand{\pmetricdp}{\textsc{Planar Metric Dimension}}

\newcommand{\npsat}{\textsc{$1$-Negative Planar 3-SAT}}
\newcommand{\sat}{\textsc{$3$-SAT}}
\newcommand{\hz}{\hat{z}}

\begin{document}

\title{Complexity of Metric Dimension \mbox{on Planar Graphs}\thanks{An extended abstract of this paper appeared as \emph{On the Complexity of Metric Dimension} in L.~Epstein, P.~Ferragina (eds.). Algorithms -- ESA 2012, 20th Annual European Symposium, Ljubljana, Slovenia, September 10-12, 2012, Proceedings. LNCS vol.~7501, Springer, 2012, pp.~419--430.}}

\author{Josep Diaz \footnote{Ci{\`e}ncies de la Computaci{\'o}, UPC, Jordi Girona Salgado 1--3, 
08034 Barcelona, Spain. Partially supported by Ministerio de Economia y Competitividad under grant TIN2013-46181-C2-1-R (COMMAS) and Generalitat de Catalunya, Ag{\`e}ncia de Gesti{\'o} d'Ajuts Universitaris i de Recerca, under project 2014 SGR 1034 (ALBCOM-RG). \texttt{\{diaz|mjserna\}@cs.upc.edu}}
\and 
Olli Pottonen \footnote{(Present address) SilverRail Australia Pty Ltd, Brisbane, Australia. Supported by the Finnish Cultural Foundation. \texttt{{olli.pottonen@iki.fi}}}$^{\ddagger,\, \dagger}$\stepcounter{footnote}
\and
Maria Serna \footnotemark[2]
\and 
Erik Jan van Leeuwen \footnote{Max-Planck Institut f\"{u}r Informatik, Campus E1 4, 66123 Saarbr\"{u}cken, Germany. \texttt{erikjan@mpi-inf.mpg.de}}}

\date{}

\maketitle

\begin{abstract}
The metric dimension of a graph $G$ is the size of a smallest subset $L \subseteq V(G)$ such that for any $x,y \in V(G)$ with $x\not= y$ there is a $z \in L$ such that the graph distance between $x$ and $z$ differs from the graph distance between $y$ and $z$. Even though this notion has been part of the literature for almost 40 years, prior to our work the computational complexity of determining the {\metricd} of a graph was still very unclear. 
In this paper, we show tight complexity boundaries for the {\metricdp} problem. We achieve this by giving two complementary results. First, we show that the {\metricdp} problem on planar graphs of maximum degree~$6$ is NP-complete. Then, we give a polynomial-time algorithm for determining the metric dimension of outerplanar graphs.
\end{abstract}

\section{Introduction} \label{intro}
In this paper, we study the complexity of the \metricdp{} problem, in particular on planar graphs. To define the \metricdp{} problem, we need several supporting notions. Let $G$ be a graph. We say that $z \in V(G)$ \emph{resolves} two vertices $x,y \in V(G)$ with $x\not= y$ if the length of a shortest path in $G$ from $z$ to $x$ is different from the length of a shortest path in $G$ from $z$ to $y$. Then a set $L \subseteq V(G)$ is called a \emph{resolving set} (or \emph{metric generator}) of $G$ if every pair $x,y \in V(G)$ with $x\not= y$ is resolved by some $z \in L$. We sometimes refer to the elements of a resolving set (or in fact, of any set of vertices that we hope to extend to a resolving set) as \emph{landmarks}. Now the \emph{\metricd} of $G$ is the cardinality of a smallest resolving set of $G$ (such a smallest resolving set is known as a \emph{metric basis}). The problem of determining the \metricd{} of a given graph $G$ is called \emph{\metricdp}, but is also known as \emph{Harary's problem} or the \emph{rigidity problem}. The problem was defined independently by Harary and Melter~\cite{HararyM76} and Slater~\cite{Slater}.

There are several reasons for studying the \metricdp{} problem. The first reason is that, even though the problem is part of Garey and Johnson's  book on computational intractability~\cite{GareyJ1979}, very little is known about the computational complexity of this problem. Garey and Johnson proved thirty years ago that the decision version of {\metricdp} is NP-complete on general graphs~\cite{David} (another proof appears in~\cite{Khuller96}). Also it was shown that there exists a $2\log n$-approximation algorithm on arbitrary graphs~\cite{Khuller96}, which is best possible within a constant factor under reasonable complexity assumptions~\cite{Beerliova06,Hauptmann11}. Hauptmann~\etal\cite{Hauptmann11} showed hardness of approximation on sparse graphs and on complements of sparse graphs. On the positive side, fifteen years ago, Khuller~\etal\cite{Khuller96} gave a linear-time algorithm to compute the {\metricd} of a tree (see also~\cite{Slater,HararyM76}), as well as a characterization for graphs with {\metricd} $1$ and several interesting properties of graphs with {\metricd} $2$. Similar results were independently obtained by Chartrand~\etal\cite{Chartrand00}.
Before we published a preprint of our work, no further results were known about the complexity of this problem. It is thus interesting if the substantial, long-standing gap on the tractability of this problem (between trees and general graphs) can be bridged.

After a preprint of our work appeared, a large number of papers have appeared that further investigate the complexity of {\metricdp} on graph classes. 
On the negative side, Epstein~\etal\cite{EpsteinLW2012} provided NP-hardness results for split graphs, bipartite graphs, co-bipartite graphs, and line graphs of bipartite graphs. Hoffman and Wanke~\cite{HoffmanW2012}, based on the NP-hardness reduction for planar graphs given in this paper, were able to prove that the problem is NP-hard on Gabriel unit disk graphs. More recently, Foucaud~\etal\cite{FoucaudMNPV2015,FoucaudMNPV2015b} showed that the problem is NP-hard on permutation graphs and interval graphs.
Fernau and Rodr{\'{\i}}guez{-}Vel{\'{a}}zquez~\cite{FernauR2014} showed that on general graphs there is no algorithm running in $O( |V(G)|^{O(1)}\, 2^{o(|V(G)|)} )$ time unless the Exponential Time Hypothesis fails; this complements their algorithm running in $O( |V(G)|^{O(1)}\, 2^{|V(G)|} )$ time.
Hartung and Nichterlein~\cite{HartungN} settled the parameterized complexity for the standard parameter (the size of the resolving set) on general graphs, by showing that the problem is W[2]-complete even if the maximum degree is at most three; they also give a strong approximation hardness result on such graphs.

On the positive side, Epstein~\etal\cite{EpsteinLW2012} presented polynomial-time algorithms for a weighted variant of {\metricdp} on several graphs including paths, trees, and cographs. Fernau~\etal\cite{FernauHHMS2015} gave a polynomial-time algorithm for {\metricdp} on chain graphs, a subclass of bipartite graphs.
Foucaud~\etal\cite{FoucaudMNPV2015,FoucaudMNPV2015b} showed that {\metricdp} is fixed-parameter tractable for the standard parameter on interval graphs. Belmonte~\etal\cite{BelmonteFGR2016} generalized this result to graphs of bounded treelength, which include not only interval graphs, but also chordal graphs, permutation graphs, and AT-free graphs.

The second reason for studying \metricdp{} is that the problem has received a lot of attention from researchers in  different disciplines, in particular as a difficult graph theoretical  problem (see e.g.~\cite{Bailey11,Caceres07,Chartrand00,Hauptmann11} and references therein). For instance, a recent survey by Bailey and Cameron~\cite{Bailey11} notes an interesting connection to group theory and graph isomorphism. It was also shown to be applicable to certain cop-and-robber games~\cite{West} and to routing in networks~\cite{Fonseca}. Therefore it makes sense to continue the investigation on the computational complexity of {\metricdp} and narrow the above-mentioned complexity gap.

The third reason for studying {\metricdp}, particularly on planar and outerplanar graphs, is that known techniques in the area do not seem to apply to it. Crucially, it seems difficult to formulate the problem as an MSOL-formula, without which we cannot apply Courcelle's Theorem~\cite{Courcelle} on graphs of bounded treewidth. Hence, there is no easy way to show that the problem is polynomial-time solvable on graphs of bounded treewidth. Also, the line of research pioneered by Baker~\cite{Baker}, which culminated in the recent meta-theorems on planar graphs using the framework of bidimensionality~\cite{DemaineH,FominLS}, does not apply, as {\metricdp} does not exhibit the required behavior. For example, the {\metricd} of a (two-dimensional) grid is two~\cite{Khuller96} (see also~\cite{Caceres07}), whereas bidimensionality requires it to be roughly linear in the size of the grid. Moreover, the problem is not closed under contraction. This behavior of {\metricdp} contrasts that of many other problems, even that of other nonlocal problems such as \textsc{Feedback Vertex Set}. Hence, by studying the {\metricdp} problem, there is an opportunity to extend the toolkit that is available to us on planar graphs.

\paragraph{Our Results}
In the present work, we significantly narrow the tractability gap of {\metricdp}. From the hardness side, we show that {\metricdp} on planar graphs, called {\pmetricdp}, is NP-hard, even for planar graphs of maximum degree~$6$. From the algorithmic side, we show that there is a polynomial-time algorithm to find the {\metricd} of outerplanar graphs. 

The crux to both of these results is our ability to deal with the fact that the {\metricdp} problem is extremely \emph{nonlocal}. In particular, a landmark can resolve vertices that are very far away from it. The paper thus focusses on constraining the effects of a landmark to a small area. The NP-hardness proof does this by constructing a specific family of planar graphs for which  {\metricdp} is essentially a local problem. The algorithm on outerplanar graphs uses a tree structure to traverse the graph, together with several data structures that track the influence of landmarks on other vertices. As we show later, this is sufficient to keep the nonlocality of the problem in check.
We believe that our algorithmic techniques are of independent interest, and could lead to (new) algorithms for a broad class of nonlocal problems.

\paragraph{Overview of the NP-Hardness Proof}
As a corollary of the work by Dahlhaus \etal\cite{DahlhausJPSY1994},
we prove a new version of {\sc Planar 3-SAT} to be NP-complete. We reduce this problem
to {\metricdp}. This is done by constructing a planar
graph consisting of clause gadgets and variable gadgets. Let $n$ be
the number of variables. Each variable gadget must have four landmarks: three
at known, specific locations, but for the fourth we have three
different choices. They correspond to the variable being true, false,
or undefined. These $4n$ landmarks are a resolving set if and only if
they resolve all pairs of vertices in the clause gadgets, which happens only if they
correspond to a satisfying truth assignment of the SAT-instance.

\paragraph{Overview of the Algorithm}
Observe that the standard dynamic-programming approach using a tree decomposition fails here, as the amount of information one needs to maintain seems to depend exponentially on $n$, rather than on the width of the decomposition. To overcome this fact we take a different approach. 

We characterize resolving sets in outerplanar graphs by giving two necessary and sufficient requirements for an arbitrary set of vertices to be a resolving set.
Then, taking as a base the duals of the biconnected components of the graph $G$, we define a tree $T$. Vertices of $T$ correspond to faces and cut vertices of $G$, and edges of $T$ correspond to inner edges and bridges of $G$.
Note that each vertex and edge of $T$ corresponds to a separator of $G$. 
The algorithm uses dynamic programming to process $T$, starting at the leaves and advancing towards the root. 

At first sight, this decomposition has the same problem as we had with tree decompositions. Moreover, the size of a face might be arbitrarily big, leading to a decomposition of arbitrary `width'. To overcome these obstacles, we introduce two data structures, called \emph{boundary conditions} and \emph{configurations}.  
\begin{itemize}
\item Boundary conditions track the effects of landmarks placed in the already processed part of the graph and the possible effects of sets of landmarks to be placed in the unexplored parts of the graphs.
\item Configurations represent the main novelty in our algorithm. Configurations control the process of combining the boundary conditions on edges towards children of the current vertex $v' \in V(T)$ into a boundary condition on the edge towards the parent of $v'$. The configurations depend on the vertices of $G$ represented by $v'$. Even though the number of vertices of $G$ represented by $v'$ may be unbounded, we show that the total number of relevant configurations is only polynomial.
\end{itemize}
By combining boundary conditions and configurations appropriately in a dynamic-programming procedure, we finally arrive at a polynomial-time algorithm.

The use of configurations presents a stark contrast with the techniques used in bounded treewidth algorithms, where the combination process commonly is a simple static procedure. A similar contrast is apparent in our tree structure: whereas outerplanar graphs have constant treewidth~\cite{Bodlaender86}, the tree structure used in our approach actually leads to a decomposition that can have arbitrary width.

\section{Preliminaries}
For basic notions and results in graph theory, we refer the reader to
any textbook on the topic, e.g.~Diestel~\cite{Diestel2000}. All graphs
are finite, undirected, and unless otherwise stated, connected. The vertex and edge sets of a graph $G$ are denoted by $V(G)$ and $E(G)$, respectively. We use the notation $\edge{u}{v}$ to denote an edge from $u$ to $v$. Given $v\in V(G)$, $\Nei (v)$ denotes the set of neighbors  of $v$ in $G$. The graph distance between vertices $u$ and $v$ is denoted by $d(u, v)$.

A graph $G$ has a {\em cut vertex} if the removal of that vertex disconnects the graph into at least two components. A graph is  a {\em biconnected} if it has no cut vertices.

A \emph{planar embedding} of a graph $G$ is an assignment of $V(G)$ to distinct points in the plane and $E(G)$ to Jordan curves (\ie simple closed curves in the plane) such that the curve of each edge starts and ends at the points corresponding to the endpoints of the edge, and no interior point on the curve is a point corresponding to a vertex of $G$ nor on a curve corresponding to another edge. In other words, the vertices are points in the plane and the edges are drawn between the points so that they do not intersect. A graph is \emph{planar} if it has a planar embedding. Equivalently, a graph is  {\em planar} if and only if it does not contain a subgraph that is a subdivision of $K_5$ or $K_{3,3}$.

An \emph{outerplanar embedding} of a graph $G$ is a planar embedding where all points corresponding to the vertices of $G$ border the outer (infinite) face. We call the edges on the border of the outer face the \emph{outer edges} of $G$, and we call the other edges \emph{inner edges}. Note that for each biconnected component $B$ of $G$, the outer edges of $B$ form a Hamiltonian cycle of $B$.
A graph $G$ is \emph{outerplanar} if it has an outerplanar embedding. Equivalently, a graph is {\em outerplanar} if and only if it does not contain a subgraph that is a subdivision of $K_4$ or $K_{2,3}$. If an outerplanar graph is given together with such an embedding, it is called \emph{outerplane}.

We also repeatedly use the following observation about outerplanar graphs. Given an outerplanar graph $G$ and a cycle $C$ of $G$, call a path \emph{$C$-disjoint} if no vertex of the path (except possibly its ends) belongs to $C$.

\begin{proposition} \label{prp:outerplanar}
Let $G$ be an outerplanar graph, let $C$ be a cycle of $G$, and let $u \in V(G) \setminus C$. Then no three distinct vertices of $C$ have $C$-disjoint paths to $u$. Moreover, any two distinct vertices of $C$ having $C$-disjoint paths to $u$ must be neighbors on $C$.
\end{proposition}
\begin{proof}
If three distinct vertices of $C$ each have a $C$-disjoint path to $u$, then these paths together with $C$ contain a subgraph that is a subdivision of $K_{4}$. This contradicts the outerplanarity of $G$. If two distinct vertices of $C$ that are not neighbors on $C$ each have a $C$-disjoint path to $u$, then these paths together with $C$ contain a subgraph that is a subdivision of $K_{2,3}$. This contradicts the outerplanarity of $G$.
\qed\end{proof}

In an outerplane graph $G$, any cycle of $G$ corresponds to a Jordan curve. 
Given two cycles $C,C'$ of $G$, we say that $C$ is \emph{topologically contained} in $C'$ if the Jordan curve of $C$ is contained in (the closure of) the interior of the Jordan curve of $C'$. Note that this is actually equivalent to stating that $V(C) \subseteq V(C')$, but the topological definition might be more intuitive.

Finally, given a set $S$, we denote by $\mathcal{P}^{k}(S)$ the set of all subsets of $S$ that have at most $k$ elements.

\section{NP-Hardness on Planar Graphs} \label{sec:hardness}
We reduce from a variation of the {\sat} problem. We first require some notation.

\begin{definition}
Let $\Psi$ be a boolean formula on a set $V$ of variables and a set $C$ of clauses. The \emph{clause-variable graph} of $\Psi$ is defined as $G_{\Psi} = (V \cup C, E)$, where $E = \{ (v,c) \mid v \in V, c \in C, v \in c \}$.
\end{definition}

\noindent
The notation $v \in c$ means that variable $v$ (or its negation) occurs in clause $C$. Observe that $G_{\Psi}$ is always bipartite.

\begin{theorem}[{\cite[p.~877]{DahlhausJPSY1994}}] \label{thm:dahlhaus}
The problem of deciding whether a boolean formula $\Psi$ is satisfiable is NP-complete, even if
\begin{itemize}
\item every variable occurs in exactly three clauses (twice positive, once negative),
\item every clause contains two or three distinct variables, and
\item $G_{\Psi}$ is planar.
\end{itemize}
\end{theorem}
As a corollary of Theorem~\ref{thm:dahlhaus}, we get the following result,
which is the starting point of our work.

\begin{corollary}\label{thm:planar3sat}
The problem of deciding whether a boolean formula $\Psi$ is satisfiable is NP-complete, even if
\begin{itemize}
\item every variable occurs exactly once negatively and once or twice
positively,
\item every clause contains two or three distinct variables,
\item every clause with three distinct variables contains at least one
negative literal, and
\item $G_{\Psi}$ is planar.
\end{itemize}
\end{corollary}
We call this decision problem {\npsat}.

\begin{proof}
Let $\Psi$ be a boolean formula satisfying the constraints of
Theorem~\ref{thm:dahlhaus}. By modifying $\Psi$ we will construct a
formula $\Psi'$ that fulfills all the constraints of the theorem
statement and is satisfiable if and only if $\Psi$ is
satisfiable.

We only need to eliminate those clauses containing three positive literals. Suppose that $x \vee y \vee z$ is such a clause of $\Psi$
with distinct variables $x,y,z$. Add a new variable $x'$, and
replace the original clause by the clauses
$x \vee x'$ and $\lnot x' \vee y \vee z.$ This
completes the construction of $\Psi'$. As this construction replaces
some edges of $G_\Psi$ with paths, it preserves planarity.

Given a satisfying truth assignment of $\Psi$, we get a satisfying
assignment of $\Psi'$ by setting $x' = \lnot x$. A satisfying
assignment of $\Psi'$ implies a satisfying assignment of $\Psi$. So $\Psi'$
is satisfiable if and only if $\Psi$ is. The theorem now follows straightforwardly from Theorem~\ref{thm:dahlhaus}.
\qed\end{proof}

\begin{figure}
\begin{center}
\scalebox{0.5}{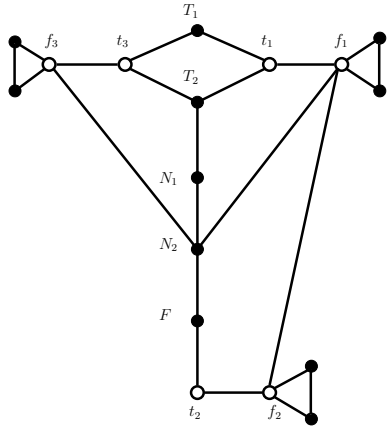 }
\end{center}
\caption{The variable gadget.}\label{fig:var}
\end{figure}
\begin{figure}
\begin{center}
\scalebox{0.5}{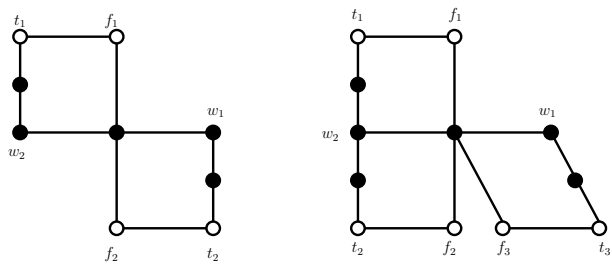 }
\end{center}
\caption{The clause gadget for a clause with two variables (left), and the gadget for a clause with three variables (right).}\label{fig:clause}
 \end{figure}
\begin{figure}
\begin{center}
\scalebox{0.45}{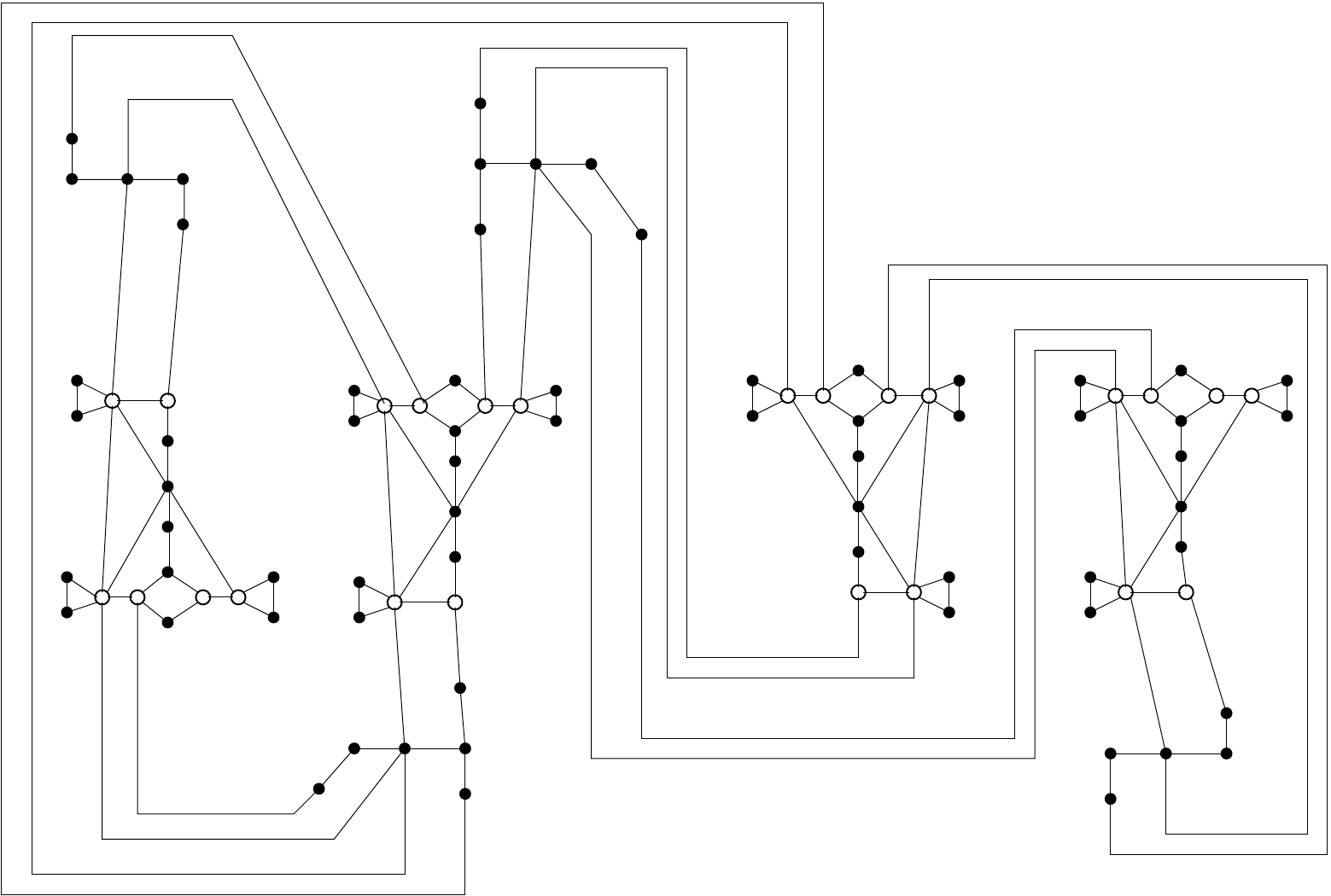 }
\end{center}
\caption{The planar graph obtained for $(\lnot{x}_1x_2)(x_1\lnot{x}_2x_3)(x_2x_4\lnot{x}_3)(\lnot{x}_4x_3)$.}\label{fig:bigone}
\end{figure}

To prove that {\pmetricdp} is NP-hard, we will give a reduction from {\npsat}. The idea behind the graph constructed in this reduction is the following. Given an instance $\Psi$ of {\npsat}, we first find a planar embedding of its clause-variable graph $G_{\Psi}$. We then replace each variable vertex of $G_{\Psi}$ by a \emph{variable gadget} (see Figure~\ref{fig:var}), and each clause vertex of $G_{\Psi}$ by a \emph{clause gadget} (see Figure~\ref{fig:clause}). By identifying vertices of variable gadgets and vertices of clause gadgets in an appropriate way (see Figure~\ref{fig:bigone}), we obtain a planar graph $H_{\Psi}$ that will be our instance of {\pmetricdp}.

We now describe our construction in detail. Consider a planar embedding of $G_{\Psi}$, which can be found in linear time~\cite{HopcroftT1974}. We first replace each variable vertex of $G_{\Psi}$ by a variable gadget. In Figure~\ref{fig:var}, the white vertices will be identified with vertices from a clause gadget later on. There are three groups (connected components) of such vertices in the figure. The groups containing vertices $(t_{1},f_{1})$ and $(t_{3},f_{3})$ will be identified with vertices in clause gadgets where this variable appears positively in the corresponding clause; the group containing $(t_{2},f_{2})$ will be identified with vertices in clause gadgets where this variable appears negatively. By rotating and contorting the variable gadget appropriately, we can ensure that the three groups point into the right direction (\ie the negative-appearance group faces the clause vertex where the variable appears negatively).

Next, we replace the clause vertices by clause gadgets. The exact gadget we use depends on whether the clause contains two or three  variables (see Figure~\ref{fig:clause}). We restrict our description to the three-variable case, as the two-variable case is similar and simpler. In Figure~\ref{fig:clause}, the white vertices will be identified with vertices from a variable gadget. There are again three groups of such vertices, one for each variable occurring in the clause.

\begin{figure}
\begin{center}
\scalebox{0.5}{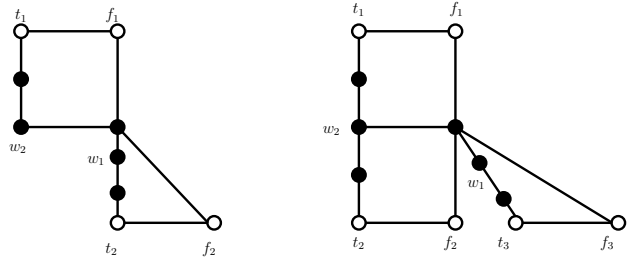 }
\end{center}
\caption{Alternative planar embeddings of the the clause gadgets.}\label{fig:case2}
\end{figure}

Obviously, we will identify the $t$-vertex of a variable group with the $t$-vertex of a clause group, and the same for the $f$-vertices. We call this \emph{matching}. It is not entirely straightforward to do this matching in a manner that preserves planarity. Consider the way in which the groups and the $t$ and $f$ vertices appear on the boundary of the clause gadget. In Figure~\ref{fig:clause}, the pairs appear in order $(t_1,f_1),(t_3,f_3),(f_2,t_2)$ clockwise starting from the top. As illustrated in Figure~\ref{fig:case2}, $(t_1,f_1),(f_3,t_3),(f_2,t_t)$ is also possible. The remaining two alternatives, $(t_1,f_1),(t_3,f_3),(t_2,f_2)$ and $(f_1,t_1),(f_3,t_3),(f_2,t_2)$ are to be avoided. This is accomplished by choosing a variable appearing negatively in the clause and mirroring the corresponding variable gadget around the axis $T_1$---$F$ (see Figure~\ref{fig:var}). This does not affect our ability to connect the variable to other clauses.

This completes the construction. Call the resulting graph $H_{\Psi}$. Observe that $H_{\Psi}$ is planar, and has maximum degree~$6$: the individual gadgets have degree at most~$5$, but in putting them together the vertex $f_1$ gets degree~$6$. 
We remark that each variable appears once negatively in $\Psi$, and once or twice positively. So if the variable appears only twice, then $(t_1, f_1)$ or $(t_3, f_3)$ in the corresponding variable gadget will not be identified with a group of vertices in a clause gadget.

In Figure~\ref{fig:bigone} we can see an example of the reduction and the resulting planar graph from the specific instance of  {\npsat} 
$(\lnot{x}_1\vee x_2)\wedge (x_1\vee \lnot{x}_2\vee x_3)\wedge (x_2\vee x_4\vee \lnot{x}_3)\wedge (\lnot{x}_4\vee x_3) $.

We now make several observations about the graph $H_{\Psi}$ and the way vertices of a resolving set need to be positioned on it.

Each $f$-vertex is contained in a triangle, say with other vertices $r,s$. Observe that $r$ and $s$ can only be resolved if $r$ or $s$ is part of the resolving set. We call these \emph{forced landmarks}. In fact, in any smallest resolving set, exactly one of $r,s$ will be a landmark. Then it follows by construction that $H_{\Psi}$ requires exactly $3n$ forced landmarks, where $n$ is the number of variables of $\Psi$.

Using the forced landmarks, we can resolve most pairs of vertices, as shown by the following lemma. We say that $T_1$, $T_2$, $N_1$, $N_2$, and $F$ are \emph{strictly inside} the variable gadget.

\begin{figure}
\begin{center}
\scalebox{0.5}{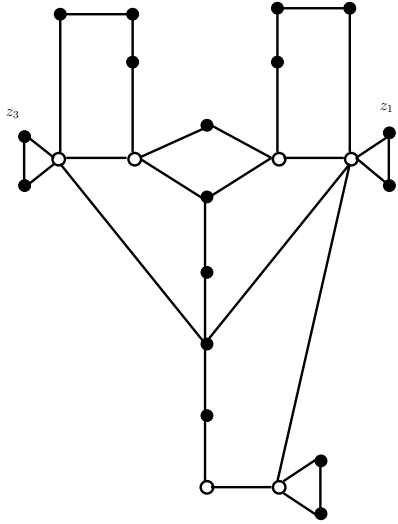 }
\end{center}
\caption{The graph in Lemma~\ref{lem:forced}.}\label{fig:z1z3}
\end{figure}

\begin{lemma}\label{lem:forced}
Let $x, y \in V(H_{\Psi})$, such that $\{x,y\}$ is not equal to $\{w_1, w_2\}$ from a single clause gadget or to $\{T_1, T_2\}$, $\{T_1, N_1\}$, or $\{T_2, N_1\}$ from a single variable gadget. Then the pair $x,y$ is resolved by a forced landmark.
\end{lemma}
\begin{proof}
A relatively easy but tedious case analysis verifies the cases where both $x$ and $y$ are in
the same clause or variable gadget. There are two remaining cases:
either $x, y$ are in different clause gadgets, or $x$ is strictly
inside a variable gadget and $y$ is outside that gadget.

Consider the first case, that is, $x,y$ are in different clause gadgets.
Denote the gadget containing $x$ by $g_{x}$, and the gadget containing $y$ by $g_{y}$. Let $z_{x}$ be a forced landmark that is closest to $x$, and let $z_{y}$ be a forced landmark that is closest to $y$. Without loss of generality, $d(x,z_{x}) \leq d(y,z_{y})$. We will show that $d(x, z_x) < d(y, z_{x})$.
Since $x, y$ are in distinct clause gadgets, for any shortest path $P$ from $y$
to $z_{x}$ there is a variable gadget $g$ such that $P$ enters $g$ via one group and leaves via another one (and the part in between is fully contained in $g$.) Let $w$ denote the vertex of both $g$ and $P$ that is closest to $y$ and let $(t,f)$ denote the corresponding group of $g$ (\ie $w=t$ or $w=f$). Let $z_f$ be the forced landmark in the triangle connected to $f$.
By the definition of $g$ and the construction of the variable gadgets, we can say the following about the edges of $P$ that appear after $w$: if $w = t$, then at least two edges of $g$ still appear plus at least one more edge (possibly also in $g$) to reach $z_x$;
if $w = f$, then at least one edge of $g$ still appears plus at least one more edge (possibly also in $g$) to reach $z_x$. 
Since $d(z_f, f) = 1$, $d(z_f, t) = 2$,
in both cases the inequality $d(w, z_f) < d(w, z_x)$ holds.
Hence, $d(x, z_x) \leq d(y, z_y) \leq d(y, z_f) \leq
d(y, w) + d(w, z_f) < d(y, w) + d(w, z_x)
= d(y, z_x)$.

Now consider the second case, and assume that $x$ is strictly inside the variable gadget part of the graph in Figure~\ref{fig:z1z3}. If $y$ is in the graph of Figure~\ref{fig:z1z3}, it can be readily verified that $x$ and $y$ are resolved by the forced landmarks of the variable gadget. 
We claim that if $y$ is outside of the picture, then $d(z_1, y) + d(z_3, y) \geq 7$, where $z_{1}$ and $z_{3}$ are the forced landmarks in the triangles attached to $f_{1}$ and $f_{3}$ respectively. This implies that $z_{1}$ or $z_{3}$ is at distance at least four from $y$, whereas the distance of $z_{1}$ and $z_{3}$ to $x$ is at most three, implying that $x$ and $y$ are resolved.

To prove the claim, note that if shortest paths from $y$ to $z_1$ and $z_3$ both contain $f_1$, then $d(z_1, y) + d(z_3, y) = d(z_1, f_1) + d(z_3, f_1) + 2d(f_1, y) = 4 + 2d(f_1, y) > 6$. The same inequality holds when shortest paths from $y$ to $z_1$ and $z_3$ both contain $f_3$. Now consider the case where shortest paths from $y$ to $z_1$ and to $z_3$ contain $f_1$ and $f_3$ respectively.
Since $y$ is not in the picture, a shortest path from $f_3$ to $y$ has at least two edges.
If a shortest path from $f_1$ to $y$ goes through the bottom group of the variable gadget in the graph of Figure~\ref{fig:z1z3}, then it has at least three edges. Otherwise, it only has two edges if $y$ is a neighbor of the top-left vertex of the left (partial) clause gadget of the graph in Figure~\ref{fig:z1z3}, but then a shortest path from $f_3$ to $y$ has at least three edges.
This gives
$d(y, f_1) + d(y, f_3) \geq 5$, and $d(z_1, y) + d(z_3, y) \geq 7$.
\qed\end{proof}

It remains to analyze how the pairs excluded in Lemma~\ref{lem:forced} can be resolved. This will rely on the satisfiability of $\Psi$, as described below, but the following auxiliary lemma is crucial.

\begin{lemma}\label{lem:variablelandmark}
All pairs of vertices that are strictly inside a variable gadget are resolved
if and only if there is a landmark strictly inside the variable gadget.
\end{lemma}
\begin{proof}
It is easy to check that a landmark that is strictly inside a variable gadget
together with the forced landmarks resolves all pairs of vertices that are strictly inside
the gadget. If no landmark is strictly inside the variable gadget, then
from any landmark $z$ there are shortest paths to $T_{1}$ and $T_2$ that both contain $t_1$ or $t_3$.
But then $d(z, T_1) = d(z, T_2)$.
\qed\end{proof}

This lemma and the forced landmarks together imply that $H_{\Psi}$ has metric dimension at least $4n$. With this fact in mind, we present the proof of the NP-hardness result.

\begin{theorem}
{\pmetricdp} is NP-complete, even on graphs of maximum degree~$6$.
\end{theorem}
\begin{proof}
Let $\Psi$ be an instance of {\npsat} with $n$ variables. Construct the graph $H_{\Psi}$ in the manner described before. Constructing $H_{\Psi}$ clearly takes time polynomial in the number of variables and clauses of $\Psi$.

We now claim that $H_{\Psi}$ has {\metricd} at most $4n$ if and only if $\Psi$ is satisfiable. Suppose that a satisfying truth assignment for $\Psi$ is given. Place the $3n$ forced landmarks. If a variable has value true, place a
landmark on $T_1$ in the corresponding gadget; otherwise, place a
landmark on $F$. After applying Lemma~\ref{lem:forced} and~\ref{lem:variablelandmark},
we only need to check that pairs $w_1, w_2$ in clause gadgets are resolved.
But each such pair is resolved by the landmark strictly inside
the variable that satisfies the corresponding clause. Hence $H_{\Psi}$ has {\metricd} at most $4n$.

Conversely, suppose that $H_{\Psi}$ has a resolving set of size $4n$. We will construct a satisfying
assignment for $\Psi$. Each variable gadget contains exactly one landmark, which is on $T_i$, $N_i$, or $F$.
If the landmark is on $T_i$, set the variable to true.
If the landmark is on $F$, set it to false. Otherwise
the variable can be arbitrarily set to either true or false. It remains to show that because the pairs $w_1, w_2$
are resolved, the truth assignment is satisfying. Note that a landmark $z$ resolves pair $w_1, w_2$ if a shortest path from a landmark to either of them enters the clause gadget through some $t_i$. 
Observe that a shortest path between $z$ and (say) $w_{1}$ that enters some (clause or variable) gadget through an $f$-vertex, by construction, will also leave that gadget through an $f$-vertex (if it leaves the gadget at all).
Hence, if a shortest path from landmark $z$ to $w_{1}$ or $w_{2}$ intersects more than
one clause gadget, it leaves the first clause through an $f$-vertex, after which
it enters all subsequent ones through an $f$-vertex. But then $w_1, w_2$ in
the final clause are not resolved. It follows that a landmark $z$ resolves $w_{1}$ and $w_{2}$ only if it is in an adjacent variable gadget and the corresponding variable satisfies the
corresponding clause. This proves the claim.

Following the claim, the reduction should construct $H_{\Psi}$ as the graph for the instance of {\pmetricdp} and set $k$ to $4n$.
\qed\end{proof}

\section{Characterizing Resolving Sets of Outerplanar Graphs} \label{sec:algorithm:charac}
In this section, we characterize resolving sets of outerplanar graphs in a way that lends itself to algorithmization. We present several intermediate results before giving the final characterization. First, we give a characterization of resolving sets in trees by giving a necessary and sufficient requirement for a set of vertices to be a resolving set. Making the step from trees to outerplanar graphs requires a closer look at a certain type of cycles, called \emph{implied cycles}. We then give a sufficient requirement for a set of vertices to be a resolving set with respect to such cycles. By generalizing this requirement, we end up with two requirements (the one for trees and a new one) that will be necessary and sufficient for a set of vertices to be a resolving set of an outerplanar graph.

Throughout the remainder of the paper, we may assume that each graph that we consider is a connected graph, as a resolving set of a disconnected graph is the union of resolving sets of its components\footnote{With one exception: isolated vertices. An edgeless graph of $n$ vertices has metric dimension $n-1$.}. We can also assume that each graph has at least three vertices; otherwise, determining the metric dimension is trivial.

We start by giving some definitions that will be used throughout the paper.

\begin{definition}
Let $G$ be a graph. A \emph{bifurcation point} associated with $z, x, y \in V(G)$ is a vertex $v \in V(G)$ farthest from $z$ such that $v$ is on shortest paths from $z$ to both $x$ and $y$. More formally, $v$ is a bifurcation point if it is on shortest paths $z \leadsto x$, $z \leadsto y$, and if any two shortest paths $v \leadsto x$, $v \leadsto y$ intersect only in $v$.
\end{definition}

Note that in an outerplanar graph the bifurcation point for each triple of vertices is unique.

As a technical trick we sometimes treat the midpoint of an inner edge $e = \edge{v_1}{v_2} \in E(G)$ as an actual vertex. The distances from this {\em midpoint vertex} $v_e$ are such that $d(v_e, v_1) = d(v_e, v_2) = \frac{1}{2}$ and  $d(v_e, x) = \min\{ d(v_e, v_1) + d(v_1, x), d(v_e, v_2) + d(v_2, x)\}$.

\begin{definition}\label{def:representative}
Let $G$ be a connected outerplanar graph with at least three vertices, let $z \in V(G)$, and let $C$ be either a single edge or a cycle. The \emph{representative} of $z$ on $C$ is the element of $V(C)$ closest to $z$, if it is unique. If it is not unique, then Proposition~\ref{prp:outerplanar} implies that there are two closest vertices, which are adjacent. In this case the representative is the midpoint of those two vertices.
\end{definition}

The cycle $C$ in Definition~\ref{def:representative} may have chords. Two kinds of cycles are especially interesting: faces and biconnected components. Note that in the latter case, the representative is never a midpoint.

We will frequently use the following result on representatives.

\begin{proposition} \label{prp:cycle:shortest}
Let $G$ be a connected outerplanar graph with at least three vertices, let $C$ be a cycle, let $z, p \in V(G)$, and let $\hz$ be the representative of $z$ on $C$. Suppose that there exists a shortest path $z \leadsto p$ that intersects $C$. If $\hz \in V(G)$, then there is a shortest path $z \leadsto p$ that contains $\hz$. Otherwise, \ie if $\hz$ is a midpoint of an edge of $C$, then there is a shortest path $z \leadsto p$ that contains an endpoint of that edge.
\end{proposition}
\begin{proof}
Let $P$ be a shortest path $z \leadsto p$ that intersects $C$. Let $v$ denote the vertex of $C$ on $P$ that is closest to $z$ on $P$. If $\hz = v$, then $P$ satisfies the lemma. So assume otherwise. Observe that the subpath of $P$ from $z$ to $v$ is a $C$-disjoint path. Since $\hz\not= v$, there is another vertex $u$ of $C$ such that $z$ has a $C$-disjoint path from $z$ to $u$. It follows from Proposition~\ref{prp:outerplanar} that the vertex $u$ is unique and that $u$ and $v$ are neighbors on $C$. If $\hz$ is the midpoint of $\edge{u}{v}$, then $P$ satisfies the lemma. If $\hz = u$, then $\dist(z,v) = \dist(z,u)+1$. Let $Q$ be the concatenation of a shortest path $z \leadsto u$, $\edge{u}{v}$, and the subpath of $P$ from $v$ to $p$. Since $\dist(z,v) = \dist(z,u)+1$, the length of $Q$ is equal to the length of $P$, and thus $Q$ is also a shortest path $z \leadsto p$. As $Q$ contains $u = \hz$, the lemma follows.
\qed\end{proof}

\subsection{A Characterization for Trees}
In this section we  provide a novel characterization of resolving sets for the case in which $G$ is a tree. We define the function $g: V(G) \times \mathcal{P}(V(G)) \rightarrow \mathcal{P}(V(G))$ as
\[g(v, L) = \{w \in \Nei (v) : d(z, w) = d(z, v) + 1 \textrm{ for all } z \in L\}. \]
In other words, a neighbor $w$ of $v$ is in $g(v, L)$ if for every $z \in L$, $v$ is on some shortest path $z \leadsto w$ (but $v$ is not necessarily on every such shortest path.)
Observe that any pair $x, y \in g(v, L)$ is left unresolved by $L$. So any resolving set $L$ satisfies the following:

\begin{Req}\label{req:g}
Any vertex $v \in V(G)$ must have $|g(v, L)| \leq 1$.
\end{Req}

\noindent
We prove that Requirement~\ref{req:g} is also sufficient if $G$ is a tree.

\begin{figure}
\begin{center}
\scalebox{0.5}{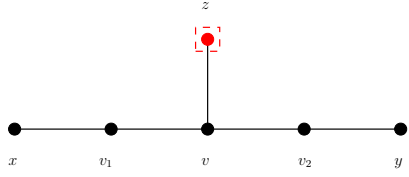 }
\end{center}
\caption{Proof of Theorem~\ref{thm:tree}.\label{fig:tree}}
\end{figure}

\begin{theorem}\label{thm:tree}
Let $G$ be a tree with at least three vertices. Then a set $L \subseteq V(G)$
is a resolving set if and only if it satisfies Requirement~\ref{req:g}.
\end{theorem}
\begin{proof}
We have already seen that any resolving set satisfies Requirement~\ref{req:g}.
Now assume that $L$ satisfies Requirement~\ref{req:g}. We pick any two
vertices $x, y \in V(G)$ and show that they are resolved.

Since $G$ has at least three vertices, there is at least one vertex
$v \in V(G)$ with degree at least $2$. Since $|g(v, L)| \leq 1 < |\Nei(v)|$,
$L$ is not empty.

Choose any $z \in L$.  If $z$ resolves $x, y$, then we are
done. Otherwise, let $v$ be the bifurcation point associated with $z, x, y$,
and let $v_1, v_2$ be the successors of $v$ on the shortest paths
$v \leadsto x, v \leadsto y$ (see Figure~\ref{fig:tree}). Since
$d(z, x) = d(z, y)$, we have $d(v, x) = d(v, y)$. By assumption, $g(v,
L)$ can not contain both $v_1$ and $v_2$. Without loss of generality
$v_1 \not \in g(v, L)$.
Then there is a vertex $z_2 \in L$ whose shortest path to $v_{1}$ does not pass through $v$. Since $G$ is a tree, the shortest path from $z_{2}$ to $v_{2}$ passes through $v_{1}$ and $v$, and thus $d(z_2,v_1) < d(z_2, v)$. As $d(v_1,x) < d(v,x) = d(v,y)$, it follows that $d(z_2, x) < d(z_2, y)$, and thus $z_{2}$ resolves $x$ and $y$.
\qed\end{proof}

As stated earlier, the major difficulty of the metric dimension problem is that it is non-local. This is why Theorem~\ref{thm:tree} is useful. Although stopping short of giving an actual local characterization of resolving sets, it does make the effects of a resolving set sufficiently local that it could be used to devise a polynomial-time algorithm for trees.

\subsection{Implied Cycles}
Our algorithm relies on a generalization of Theorem~\ref{thm:tree} to outerplanar graphs. The main difficulty in outerplanar graphs is to deal with cycles. Any resolving set needs at least two representatives on any cycle. This, fortunately, is still guaranteed by Requirement~\ref{req:g}.

\begin{lemma} \label{lem:cycle:two}
Let $G$ be a connected outerplanar graph with at least three vertices, let $C$ be a cycle of $G$, and let $L \subseteq V(G)$ satisfy Requirement~\ref{req:g}. Then $L$ has at least two representatives on $C$.
\end{lemma}
\begin{proof}
As in the proof of Theorem~\ref{thm:tree}, $L$ has to be nonempty. So there is a $z \in L$. Let $\hz$ be the representative of $z$ on $C$. We have two cases: either $\hz$ is a regular vertex, or it is a midpoint.

If $\hz$ is a vertex of $C$, then let $v_{1},v_{2}$ be the neighbors of $\hz$ on $C$. Since $\hz$ is a vertex of $C$, $\dist(z, v_1) = \dist(z,v)+1 = \dist(z,v_2)$. Hence, $|g(v,\{z\})| \geq 2$. Because $|g(v,L)|\leq 1$ by Requirement~\ref{req:g}, there is a $z' \in L \setminus\{z\}$ such that, say, $d(z', v_1) \leq d(z', \hz)$. Then the representative of $z'$ can not be $\hz$, and thus $L$ has at least two representatives on $C$.

Suppose then that $\hz$ is a midpoint of some edge $e = \edge{v_{1}}{v_{2}}$ of $C$. Let $w$ be the bifurcation point of $z,v_{1},v_{2}$. Then $d(w, v_1) = d(w, v_2)$. Denote by $s_{1}$ and $s_{2}$ the successor of $w$ on a shortest path $P_{1}$ from $w$ to $v_{1}$ and a shortest path $P_{2}$ from $w$ to $v_{2}$ respectively. Observe that $P_{1}$ and $P_{2}$ are unique and that $C \setminus\{e\}$ together with $P_{1}$ and $P_{2}$ form a cycle $C'$.

Since $w$ is the bifurcation point of $z,v_1,v_2$, the definition of $s_1,s_2$ implies that $g(w,\{z\}) \supseteq \{s_1,s_2\}$ and thus that $|g(w,\{z\})| \geq 2$. Then by Requirement~\ref{req:g}, there is a vertex $z' \in L \setminus\{z\}$ such that, say, $d(z',s_{1}) \leq d(z', w)$. Let $\hz'$ be the representative of $z'$ on $C'$.
If $\hz'$ is on $C$, then it is the representative of $z'$ on $C$. Since $e$ is not part of $C'$, $\hz \not= \hz'$, and thus $L$ has at least two representatives on $C$.
If $\hz' = w$, then $d(z', s_1) = d(z', w) + 1$, a contradiction.
Hence, \wlogen, $\hz'$ is on $P_{1}$, but is not either of its endpoints. We claim that $d(z',v_1) < d(z',v_2)$, and thus $L$ has at least two representatives on $C$. To see this, by Proposition~\ref{prp:outerplanar}, $z'$ has $C'$-disjoint paths to at most two vertices of $C'$. The position of $\hz'$ implies that all are on $P_{1}$. Let $x$ be the first vertex of $C'$ on a shortest path $z' \leadsto v_2$. If $x = w$, then as $\hz' \not= w$, there is a vertex $y$ on $P_{1}$ such that $d(z',y) \leq d(z',x)$. Since $d(y,v_1) < d(x,v_1)$, $d(z',v_1) \leq d(z',y) + d(y,v_1) < d(z',x) + d(x,v_1) = d(z',x) + d(x,v_2) = d(z',v_2)$. If $x \not= w$, then as $w$ is the bifurcation point of $z,v_1,v_2$, $d(x,v_1) < d(x,v_2)$. Hence, $d(z',v_1) \leq d(z',x) + d(x,v_1) < d(z',x) + d(x,v_2) = d(z',v_2)$. In both cases, $d(z',v_1) < d(z',v_2)$. The lemma follows.
\qed\end{proof}

Having at least two representatives on each cycle, however, is not enough to guarantee that a set $L$ of vertices is a resolving set. In fact, Requirement~\ref{req:g} is not strong enough to guarantee this. For example, if $G$ is an even cycle and $L$ consists of two antipodal vertices of the cycle, then Requirement~\ref{req:g} is satisfied, but $L$ is not a resolving set. 
Therefore, to give a characterization of resolving sets of outerplanar graphs, we need a new requirement that deals with cycles. In fact, the new requirement will deal with a special type of cycles, called \emph{implied cycles}. Consider the following lemma.

\begin{figure}[tbp]
\begin{center}
\scalebox{0.5}{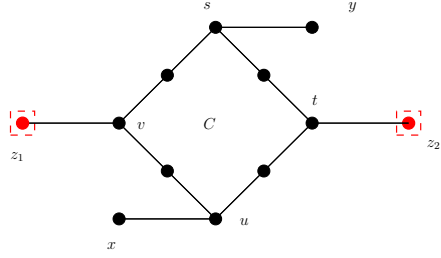 }
\end{center}
\caption{The cycle $C$ implied by $z_1, z_2, x, y$ (Lemma~\ref{lem:cycle}).}
\label{fig:spaths}
\end{figure}

\begin{lemma}\label{lem:cycle}
Let $G$ be a connected outerplanar graph and let $z_1, z_2, x, y \in V(G)$ be four distinct vertices such that neither $z_1$ nor $z_2$ resolves $x$ and $y$, and that no two shortest paths $z_1 \leadsto x$, $z_2 \leadsto y$ intersect. Then there is a cycle $C$ in $G$ such that any four shortest paths $z_1 \leadsto x, z_1 \leadsto y, z_2 \leadsto x, z_2 \leadsto y$ contain all vertices of $C$.
\end{lemma}
\begin{proof}%
We start by proving that no two shortest paths  $z_1 \leadsto y$ and $z_2 \leadsto x$ intersect. Assume to the contrary that there is a vertex $w$ that is on both a shortest path $z_1 \leadsto y$ and a shortest path $z_2 \leadsto x$. Then the assumptions and the triangle inequality yield
\begin{eqnarray*}
d(z_1, x) + d(z_2, y) & \leq & \left(d(z_1, w) + d(w, x)\right) + \left(d(z_2, w) + d(w, y)\right) \\
& = & \left(d(z_1, w) + d(w, y)\right) + \left(d(z_2, w) + d(w, x)\right) \\
& = & d(z_1, y) + d(z_2, x) \\
& = & d(z_1, x) + d(z_2, y).
\end{eqnarray*}
Therefore, $d(z_1, x) = d(z_1, w) + d(w, x)$, and $w$ is on a shortest path $z_1 \leadsto x$. Similarly, $w$ is on a shortest path $z_2 \leadsto y$.  This contradicts the assumption that no two shortest paths $z_1 \leadsto x$, $z_2 \leadsto y$ intersect.

Now let $v$ be the bifurcation point of $z_1, x, y$, and let $t$ be the bifurcation point of $z_{2},x,y$. Let $s$ denote the bifurcation point of $y,v,t$, and let $u$ denote the bifurcation point of $x,v,t$.

The vertices $v,t,s,u$ define the cycle $C$ as follows.
Let $P_{vx}$ be a shortest path $v \leadsto x$ that contains $u$. Note that $P_{vx}$ is also a shortest path $v \leadsto x$, as $u$ is the bifurcation point of $x,v,t$. Define $P_{vy}$, $P_{tx}$, and $P_{ty}$ similarly as $P_{vx}$. Finally, let $P_{vs}$ denote the subpath of $P_{vy}$ until vertex $s$, and define $P_{vu}$, $P_{ts}$, and $P_{tu}$ similarly. Figure~\ref{fig:spaths} depicts these definitions.

We claim that $P_{vs}$, $P_{vu}$, $P_{ts}$, and $P_{tu}$ are pairwise internally vertex-disjoint. This is clear for $P_{vs}, P_{vu}$ and $P_{ts}, P_{tu}$ (as $v$ and $t$ are bifurcation points) and for $P_{vs}, P_{ts}$ and $P_{vu}, P_{tu}$ (as $s$ and $u$ are bifurcation points). If $P_{vs}$ and $P_{tu}$ share a vertex, then there is a shortest path $z_1 \leadsto y$ that intersects a shortest path $z_2 \leadsto x$, a contradiction. Similarly, $P_{ts}$ and $P_{vu}$ do not share a vertex. The claim follows.

The proof of the claim actually implies that $s$, $t$, $u$, and $v$ are distinct vertices. In fact, if say $s$ and $v$ coincide, then $d(t,x) = d(t,y) = d(t,v) + d(v,y) = d(t,v) + d(v,x)$. Hence, there is a shortest path $z_{1} \leadsto y$ that intersects a shortest path $z_{2} \leadsto x$, a contradiction. Therefore, the combination of $P_{vs}$, $P_{vu}$, $P_{ts}$, and $P_{tu}$ forms a cycle $C$.

Suppose now that there exist four shortest paths $z_1 \leadsto x, z_1 \leadsto y, z_2 \leadsto x, z_2 \leadsto y$ that do not contain a vertex $w$ of $C$. Without loss of generality, we assume that $w \in P_{vs}$. Let $P$ be a shortest path $z_1 \leadsto y$ that contains $P_{vs}$ as a subpath, and let $\hat{P}$ be a shortest path $z_1 \leadsto y$ that does not contain $w$. Suppose that $w$ is an internal vertex of $P_{vs}$. Since $P$ and $\hat{P}$ are both shortest paths $z_{1} \leadsto y$ and $s$ is the bifurcation point of $y,v,t$, $\hat{P}$ cannot contain an inner edge of $C$. Then $P$, $\hat{P}$, and $C$ contain a subdivision of $K_{2,3}$ as a subgraph, contradicting the outerplanarity of $G$. So suppose that $w = s$. Let $Q$ be a shortest path $z_2 \leadsto y$ that contains $P_{ts}$ as a subpath, and let $\hat{Q}$ be a shortest path $z_2 \leadsto y$ that does not contain $s$. In a manner similar as for $\hat{P}$, we can argue that $\hat{Q}$ cannot contain an inner edge of $C$. Then $P$, $\hat{P}$, $Q$, $\hat{Q}$, and $C$ contain a subdivision of $K_{2,3}$ as a subgraph, contradicting the outerplanarity of $G$. The case that $w = v$ is similar.
\qed\end{proof}

\noindent
The lemma leads to the definition of an implied cycle.

\begin{definition} \label{def:implied}
Let $G$ be a connected outerplanar graph. Given four distinct vertices $z_1,z_2,x,y \in V(G)$ that satisfy the conditions of Lemma~\ref{lem:cycle}, the cycle $C$ of $G$ whose existence follows from Lemma~\ref{lem:cycle} is said to be \emph{implied} by $z_1,z_2,x,y$. Given a set $L \subseteq V(G)$ of vertices, we say that a cycle $C$ of $G$ is an \emph{implied cycle} of a set $L$ if $C$ is implied by some $z_1, z_2\in L$ and $x, y \in V(G)$. 
The bifurcation points $v$ (of $z_1,x,y$), $t$ (of $z_2,x,y$), $s$ (of $y,v,t$), and $y$ (of $x,v,t$) are the \emph{defining representatives} of the cycle.
\end{definition}
Note that Lemma~\ref{lem:cycle} shows that the implied cycle consists of the four defining representative and the shortest paths between them (see Figure~\ref{fig:spaths}). 

The following proposition shows that defining representatives indeed are representatives according to Definition~\ref{def:representative}.

\begin{proposition} \label{prp:cycle:repr}
Let $G$ be a connected outerplanar graph, let $x,y,z_1,z_2 \in V(G)$ be four distinct vertices, let $C$ be a cycle of $G$ implied by $z_{1},z_{2}, x, y$.
Then the defining representatives of $C$ are the representatives of $z_{1}, z_{2}, y$ and $x$ on $C$.
\end{proposition}
\begin{proof}
Let $v$ be the defining representative which is the bifurcation point of
$z_1, y, x$. We will prove that it is the representative of $z_1$ on $C$.
The three other cases are similar.

Suppose that $z_{1}$ has a shortest path to $C$ ending at a vertex $w \not= v$. Then $w$ is unique and has to be a neighbor of $v$ by Proposition~\ref{prp:outerplanar}. Assume \wloge that $w$ is on $P_{vs}$, where $P_{vs}$ is the shortest path from $v$ to $s$. Then $d(w,y) < d(v,y)$. Since $d(z_{1},w) \leq d(z_{1},v)$, it follows that $d(z_{1},w) + d(w,y) < d(z_{1},v) + d(v,y)$, contradicting that $v$ is on a shortest path $z_{1} \leadsto y$, and thus contradicting the fact that $v$ is the bifurcation point of $z_{1},x,y$. Hence, any shortest path from $z_{1}$ to $C$ ends at $v$, and $v$ is the representative of $z_{1}$ on $C$.
\qed\end{proof}

\noindent
By definition, if $L \subseteq V(G)$ is a resolving set of $G$ and $z_1, z_2\in L$, $x, y \in V(G)$ imply a cycle $C$, then $L\setminus\{z_1,z_2\}$ contains a vertex to resolve $x$ and $y$. In the next subsections, we explore the properties of such a third vertex.

First, however, we give one important property of implied cycles.

\begin{proposition} \label{prp:cycle:w}
Let $G$ be a connected outerplanar graph, let $x,y,z_1,z_2 \in V(G)$ be four distinct vertices, let $C$ be a cycle of $G$ implied by $z_{1},z_{2}, x, y$, and let $w$ be a vertex of $C$ or the midpoint of an edge of $C$. Then
$d(w, y) \neq d(w, x)$, unless $w$ is
the representative of $z_1$ or $z_2$ on $C$.
\end{proposition}
\begin{proof}
Let $v,t,s,u$ be the defining representatives of $C$. Without loss of generality, $w$ is on the shortest path from $v$ to $s$.
Note that $w \neq v$, because $v$ is the representative of $z_1$ on $C$ by Proposition~\ref{prp:cycle:repr}. Since $v$ is the bifurcation point of $z_1,x,y$, it follows that $w$ is not on a shortest path $v \leadsto x$, implying that $d(v,x) < d(v,w) + d(w,x)$.
By definition, 
$d(v,x) = d(v,y)$. Since $d(v,y) = d(v,s) + d(s,y)$ by the definition of $s$ and $d(v,s) = d(v,w) + d(w,s)$, we have that $d(v,y) = d(v,w) + d(w,s) + d(s,y) \geq d(v,w) + d(w,y)$, and thus $d(w,y) < d(w,x)$.
\qed\end{proof}

\subsection{Three Representatives}
In this section, we give a sufficient condition for a set $L$ of vertices to be a resolving set. Namely, the results of this section imply that if $L$ satisfies Requirement~\ref{req:g} and $L$ has at least three representatives on all implied cycles, then $L$ is a resolving set. Note, however, that this condition is not necessary; in particular, there are resolving sets that have only two representatives on some implied cycles. Therefore, a more complicated requirement (Requirement~\ref{req:cycle}) will be stated in Section~\ref{sec:sub:charac}. Since the case when L has three representatives on an implied cycle is an important case in the correctness proof of that requirement, we still treat this situation here.

\begin{lemma} \label{lem:implied:three}
Let $G$ be a connected outerplanar graph, let $L \subseteq V(G)$ be a set of vertices, let $z_1,z_2 \in L$ and $x,y\in V(G)$ be four distinct vertices, and let $C$ be a cycle of $G$ implied by $z_1, z_2,x, y$. If $L$ has at least three representatives on $C$, then $L$ resolves $x$ and $y$.
\begin{figure}[tbp]
\begin{center}
\scalebox{0.5}{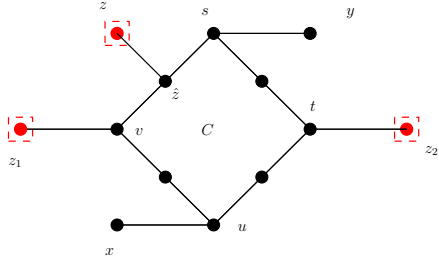 }
\end{center}
\caption{The set $L$ has three representatives on the cycle $C$ implied by $z_1, z_2, x, y$ and it resolves $x$ and $y$ (Lemma~\ref{lem:implied:three}).}
\label{fig:spaths2}
\end{figure}

\end{lemma}
\begin{proof}
Let $v, t, s, u$ be the defining representatives of $C$, and let $P_{vs}$ be the shortest path between $v$ and $s$ (see Figure~\ref{fig:spaths2}). As proved in Proposition~\ref{prp:cycle:repr}, $v$ and $t$ are the representatives on $C$ of $z_1$ and $z_2$ respectively. By assumption, there exists a $z \in L$ for which the representative $\hz$ on $C$ is different from the representatives of $z_{1}$ and $z_{2}$ on $C$ (\ie $\hz \neq v$, $\hz \neq t$). We will prove that $z$ resolves $x$ and $y$.

Without loss of generality, $\hz$ lies on $P_{vs}$. Since $\hz \not=v$, there is a $C$-disjoint path from $z$ to a vertex of $V(P_{vs}) \setminus \{v\}$. Then there is no $C$-disjoint path from $z$ to $x$, as such a path could be extended to a $C$-disjoint path from $z$ to $u$, contradicting Proposition~\ref{prp:outerplanar}. Hence, since $\hz \not= v$ lies on $P_{vs}$, Proposition~\ref{prp:cycle:shortest} implies that there exists a shortest path $z \leadsto x$ for which the vertex $r$ of $C$ that is closest to $z$ lies on $P_{vs}$.  If $r \not= v$, then Proposition~\ref{prp:cycle:w} implies that $d(r,y) < d(r,x)$, and thus $d(z, y) \leq d(z, r) + d(r, y) < d(z, r) + d(r, x) = d(z, x)$. Hence, $z$ resolves $x$ and $y$. If $r = v$, then as $\hz \not= v$, $r$ has a neighbor $r'$ on $P_{vs}$ such that $d(z, r') \leq d(z, r)$. Since $r'$ is on a shortest path $v \leadsto y$, $d(r', y) < d(r, y)$. Therefore, $d(z, y) \leq d(z, r') + d(r', y) < d(z, r) + d(r, y) = d(z, r) + d(r, x) = d(z, x)$, and thus $z$ resolves $x$ and $y$.
\qed\end{proof}

In Corollary~\ref{cor:three-rep} we prove that a similar result holds under the condition that $L$ has at least three representatives on a face of the embedding. To this end, we need the following auxiliary result.

\begin{lemma} \label{lem:three-rep}
Let $G$ be a connected outerplane graph and let $C,C'$ be cycles of $G$ such that $C'$ is topologically contained in $C$. If vertices $z_{1}\not=z_{2}$ of $G$ have the same representative $v$ on $C$, then they have the same representative on $C'$.
\end{lemma}
\begin{proof}
Suppose that $v$ is a regular vertex. 
By Proposition~\ref{prp:cycle:shortest}, any vertex of $C$ has a shortest path to $z_{1}$ that contains $v$.
In other words, $d(z_{1},c) = d(z_{1},v) + d(v,c)$ for any vertex $c$ of $C$. The same holds with respect to $z_{2}$. Since $V(C') \subseteq V(C)$, the closest vertex (or vertices) of $C'$ is the same for $z_{1}$ and $z_{2}$. Hence, $z_{1}$ and $z_{2}$ have the same representative on $C'$.

Suppose that $v$ is the midpoint of an edge $\edge{v_1}{v_2}$ of $C$. Then, by Proposition~\ref{prp:cycle:shortest}, any path from $z_1$ or $z_2$ to a vertex $c \in V(C)$ contains $v_1$ or $v_2$. In particular, $d(z_1, c) = \min\{d(z_1, v_1) + d(v_1, c), d(z_1, v_2) + d(v_2, c)\} = d(z_1, v) + d(v, c) - 1$. Using the same arguments as above, $z_{1}$ and $z_{2}$ have the same representative on $C'$.
\qed\end{proof}

\begin{corollary} \label{cor:three-rep}
Let $G$ be a connected outerplane graph and let $C,C'$ be cycles of $G$ such that $C'$ is topologically contained in $C$. For any integer $k$, if a set $L \subseteq V(G)$ of vertices has at least $k$ distinct representatives on $C'$, then $L$ has at least $k$ distinct representatives on $C$.
\end{corollary}

This corollary, together with Lemma~\ref{lem:implied:three}, implies the following.

\begin{corollary} \label{cor:implied:three:face}
Let $G$ be a connected outerplane graph, let $L \subseteq V(G)$ be a set of vertices, let $z_1,z_2 \in L$ and $x,y\in V(G)$ be four distinct vertices, let $C$ be a cycle of $G$ implied by $z_1, z_2,x, y$, and let $C'$ be a cycle of $G$ that is topologically contained in $C$. If $L$ has at least three representatives on $C'$, then $L$ resolves $x$ and $y$.
\end{corollary}

\subsection{Two Representatives and Extreme Representatives}
The previous section shows that if a set $L$ has at least three representatives on every implied cycle $C$, then any pair of vertices $x$ and $y$ are resolved by $L$. Therefore, if we know that a set of vertices has three representatives on every implied cycle, then it is not necessary to know exactly which vertices that set contains. However, if a set $L$ has two representatives on $C$, then we have to look at $L$ in more detail to determine whether it is a resolving set. In the theme of providing ``somewhat local'' characterizations, we shall prove that there is a representative that resolves $x$ and $y$ in a face $F$ ``close to'' an implied cycle $C$. In this section, we specify how to find $F$ and this representative on $F$ when $L$ and $C'$ are given.

We need some auxiliary definitions, which are standard in graph theory. We say that two faces are \emph{adjacent} if they share an edge. Note that in an outerplanar graph the vertices of such an edge actually form a separator of the graph. The \emph{weak dual}\label{def:weakdual} of an outerplane graph $G$ is a graph that has the faces of $G$ (except the unbounded outer face) as vertices. Two vertices of the weak dual are adjacent if and only if the corresponding faces are adjacent. Observe that the weak dual of a biconnected outerplane graph is a tree, and that the weak dual of an outerplane graph is a forest.

The weak dual immediately implies a distance metric on the set of faces of the outerplane graph. We need this metric in the following definition.

\begin{definition}\label{def:extremeface}
Let $G$ be a connected outerplane graph and let $L \subseteq V(G)$ have exactly one representative on a face $C'$ of $G$. An \emph{extremal face} of $L$ and $C'$ is any face $F$ of $G$ on which $L$ has at least two representatives and that is in the same biconnected component as $C'$. An \emph{extreme face} of $L$ and $C'$ is an extremal face $F$ of $L$ and $C'$ such that there is no extremal face of $L$ and $C'$ on the path in the weak dual of $G$ between $F$ and $C'$.
\end{definition}

Note that there might be a combination of a set $L$ of vertices and a face $C'$ that has no extreme faces.

We make the following observations about extreme faces.

\begin{proposition} \label{prp:extreme:atmosttwo}
Let $G$ be a connected outerplane graph and let $L \subseteq V(G)$ have exactly one representative on a face $C'$ of $G$. Then $L$ and $C'$ have at most two extreme faces.
\end{proposition}
\begin{proof}
Suppose for sake of contradiction that $L$ and $C'$ have three extreme faces $F_{1}$, $F_{2}$, and $F_{3}$. Consider the subtree of the weak dual induced by $F_{1}, F_{2}, F_{3}$ and the shortest paths between them. By the definition of an extreme face, this subtree contains a face $D$ that has degree $3$ in the subtree. Therefore, $D$ has three (inner) edges, say $f_{1}$, $f_{2}$, and $f_{3}$, such that $f_{1}$ separates $F_{1}$ from $F_{2}$ and $F_{3}$, etc. 
Since $f_{1}$, $f_{2}$, and $f_{3}$ do not have a vertex in common and $L$ can have only one representative $\hz$ on $D$, one of the inner edges (say $f_{3} = \edge{v_3}{w_3}$) is such that all vertices of $L$ are in the connected component of $G - \{v_{3},w_{3}\}$ that contains $\hz$. Note that any path from a vertex $z \in L$ to a vertex $v \in V(F_{3})$ must contain a vertex of $f_{3}$ and thus a vertex of $D$. If $\hz$ is a regular vertex, then Proposition~\ref{prp:cycle:shortest} implies that there is a shortest path from $z$ to $v$ that contains $\hz$. Hence, $L$ can have only one representative on $F_{3}$, a contradiction to the assumption that $F_{3}$ is an extreme face. If $\hz$ is the midpoint of an edge $e$, then there can be two shortest paths from $z$ to $v$ that each contain a different endpoint of $e$. Still, $L$ can have only one representative on $F_{3}$, a contradiction to the assumption that $F_{3}$ is an extreme face.
\qed\end{proof}

\begin{proposition} \label{prp:extreme1}
Let $G$ be a connected outerplane graph, let $L \subseteq V(G)$ have exactly one representative $\hz$ on a face $C'$ of $G$, and let $F$ be an extreme face of $L$ and $C'$. Then there is a unique representative of $L$ on $F$ that is furthest from $\hz$.
\end{proposition}
\begin{proof}
Let $F'$ denote the face adjacent to $F$ on the path in the weak dual between $F$ and $C'$ (note that possibly $F' = C'$), and let $e = \edge{v_1}{v_2}$ denote the edge that $F$ and $F'$ share. Since $F$ is an extreme face, $F'$ cannot be extremal, and thus $L$ has exactly one representative $\hz'$ on $F'$. This implies that all vertices of $L - \{v_1,v_2\}$ are in the component of $G - \{v_1,v_2\}$ that contains the vertices of $V(F-\{v_1,v_2\})$ (we say that $L$ is \emph{on the side of $F$}).

Suppose that $\hz'$ is the midpoint of $e$. Then the only representative of $L$ on $F$ is the vertex or midpoint that lies antipodal to $\hz'$ on $F$. This contradicts the assumption that $F$ is an extreme face (and thus in particular that $L$ must have at least two representatives on $F$).
Hence, $\hz'$ is not the midpoint of $e$. Without loss of generality, $\hz'=v_1$. Because $L$ is on the side of $F$ and $\hz'=v_1$, every representative of $L$ on $F$ is (strictly) closer to $v_1$ than to $v_2$. Since $e$ separates $C'$ and $F$, any shortest path from $\hz$ to a vertex of $F$ must contain $v_1$ or $v_2$, and thus there is always a shortest path from $\hz$ to any representative of $L$ on $F$ that contains $v_1$. It follows that a representative of $L$ on $F$ that is furthest from $\hz$ is the unique representative of $L$ on $F$ that is furthest from $v_1$ (note that this representative is strictly closer to $v_1$ than to $v_2$).
\qed\end{proof}

\begin{proposition} \label{prp:extreme2}
Let $G$ be a connected outerplane graph and let $L \subseteq V(G)$ have exactly one representative $\hz$ on a face $C'$ of $G$. If $L$ and $C'$ do not have an extreme face, then $L$ has only one representative on the biconnected component that contains $C'$.
\end{proposition}
\begin{proof}
We prove the contrapositive: if $L$ has two representatives on a biconnected component $X$, then $X$ contains an extreme face. In fact, it suffices to prove that it contains an extremal face.

Let $z, z' \in L$ have distinct distinct representatives $\hat{z},
\hat{z}'$ on $X$. Then any shortest path from $z$ to $z'$ contains
$\hat{z}, \hat{z}'$, and it contains at least one edge $e$ of
$X$. Let $F$ be a face that contains $e$.

If $z, z'$ have the same representative on $F$, then the same vertex
of $F$ has minimal distance to both $z, z'$. This is not possible,
since it contradicts the fact that $e$ is on a shortest path from $z$
to $z'$. Therefore, $L$ has at least two representatives on $F$, and it
is extremal.
\qed\end{proof}

Using extreme faces and the above propositions, we can define so-called single-extreme representatives.

\begin{definition}\label{def:extremerepr1}
Let $G$ be a connected outerplane graph and let $L \subseteq V(G)$ have exactly one representative $\hz$ on a face $C'$ of $G$. Then a \emph{single-extreme representative} $v$ of $L$ and $C'$ is the representative of $L$ farthest from $\hz$ on an extreme face of $L$ and $C'$, or if $L$ and $C'$ do not have an extreme face, then $v$ is the single representative of $L$ on the biconnected component that contains $C'$.
\end{definition}

\begin{figure}[tbp]
\begin{center}
\scalebox{0.5}{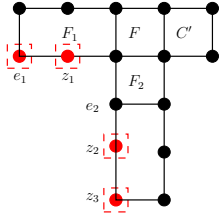 }
\end{center}
\caption{An example of single-extreme representatives. The extreme faces of $L = \{e_1, z_1, z_2, z_3\}$ and the face $C'$ are $F_{1}$ and $F_{2}$: $L$ has one representative on $F$ but several on both $F_1$ and $F_2$. The single-extreme representatives of $L$ and $C'$ are $e_1$ and $e_2$ (Definition~\ref{def:extremerepr1}). Note that $e_1 \in L$ but $e_2 \not \in L$.}
\label{fig:extreme}
\end{figure}

It follows from Proposition~\ref{prp:extreme1} and~\ref{prp:extreme2} that single-extreme representatives are well defined. Figure~\ref{fig:extreme} contains an example of single-extreme representatives.
Observe that Definition~\ref{def:extremerepr1}, in conjunction with Proposition~\ref{prp:extreme:atmosttwo}, \ref{prp:extreme1} and~\ref{prp:extreme2}, implies that $L$ and $C'$ have at least one and at most two single-extreme representatives.

We can extend Definition~\ref{def:extremerepr1} to the case when $L$ has two representatives on a face $C'$.

\begin{definition}\label{def:extremerepr}
Let $G$ be a connected outerplane graph and let $L \subseteq V(G)$ have exactly two representatives $\hz_{1},\hz_2$ on a face $C'$ of $G$.
Then $L$ can be partitioned into two sets, $L = L_1 \uplus L_2$, such that $L_{1}$ has exactly representative $\hz_1$ on $C'$ and $L_{2}$ has exactly representative $\hz_2$ on $C'$. The \emph{extreme representatives} of $L$ and $C'$ are the single-extreme representatives of $L_1$ and $C'$, and of $L_2$ and $C'$.
\end{definition}

With this definition, $L$ and $C'$ have at least one and at most four extreme representatives.

\begin{figure}[tbp]
\begin{center}
\scalebox{0.5}{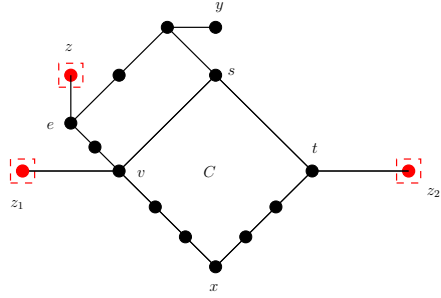 }
\end{center}
\caption{The set $L = \{z_1,z_2,z\}$ has two representatives on the cycle $C$ implied by $z_1, z_2, x, y$. Neither $z_1$ nor~$z_2$ resolves the pair $x,y$, but there is an extreme representative of $L$ and $C$ that does resolve the pair, namely~$e$, cf.~Lemma~\ref{lem:cycle:neighbors} and Requirement~\ref{req:cycle} for $C'=C$.}
\label{fig:extremeres}
\end{figure}

Extreme representatives play a crucial role in detecting whether the vertices $x$ and $y$ of an implied cycle are resolved, as we show in the following lemma. For an example of the lemma, see Figure~\ref{fig:extremeres}.

\begin{lemma} \label{lem:cycle:neighbors}
Let $G$ be a connected outerplane graph, let $L \subseteq V(G)$ be a set of vertices, let $z_1, z_2 \in L$ and $x, y \in V(G)$ be four distinct vertices that imply a cycle $C$ of $G$ such that $L$ has exactly two representatives on $C$, and let $C'$ be any face that is topologically contained in $C$. If $L$ is a resolving set, then one of the extreme representatives of $L$ and $C'$ resolves $x$ and $y$.
\end{lemma}
\begin{proof}
Let $v, t, s, u$ be the defining representatives of $C$, and define $L_1, L_2$ as in Definition~\ref{def:extremerepr} such that $z_1 \in L_1$. Since $L$ is a resolving set, there is a $z \in L$ that resolves $x$ and $y$. Without loss of generality, $z$ has $v$ as its representative on $C$ (and thus $z \in L_{1}$) and $d(z,y) < d(z,x)$. We will prove that $\edge{v}{s} \in E(G)$, that this edge separates the interior of $C$ from an extreme face $F$ of $L_{1}$ and $C'$, and that the extreme representative of $L_{1}$ and $C'$ with respect to $F$ resolves $x$ and $y$.

We start by proving that $\edge{v}{s} \in E(G)$.
If a shortest path $z \leadsto y$ intersects $C$, then because $v$ is the representative of $z$ on $C$, Proposition~\ref{prp:cycle:shortest} implies that there is a shortest path $z \leadsto y$ that contains $v$, and thus $d(z, y) = d(z, v) + d(v, y) = d(z, v) + d(v, x) \geq d(z, x)$, a contradiction. Therefore, no shortest path $z \leadsto y$ intersects $C$. Then a shortest path $z \leadsto y$ can be extended to a $C$-disjoint path $z \leadsto s$. As $v$ is the representative of $z$ on $C$, there is also a $C$-disjoint path $z \leadsto v$. Proposition~\ref{prp:outerplanar} then implies that $v$ and $s$ are adjacent. 

Observe that the above paragraph implies that there is a face that is separated from the interior of  $C$ by $\edge{v}{s}$. Call this face $F$. We claim that $F$ is an extreme face of $L_{1}$ and $C'$ and that the extreme representative of $L_{1}$ and $C'$ with respect to $F$ resolves $x$ and $y$. 

As an intermediate result, we prove that the representative $\hz$ of $z$ on $F$ resolves $x$ and $y$. Observe first that any path $z \leadsto x$ contains $v$ or $s$; otherwise, there would exist $C$-disjoint paths from $z$ to $v$, to $s$, and to a third vertex of $C$, contradicting Proposition~\ref{prp:outerplanar}. Also, since $F$ contains $v$ and $s$, any path $z \leadsto x$ intersects $F$.

Suppose that no shortest path $z \leadsto y$ intersects $F$. Let $w$ be the last vertex of $F$ on a shortest path $P_{vy}$ from $v$ to $y$ that contains $s$. Observe that in the union of the assumed $F$-disjoint path $z \leadsto y$ and $P_{vy}$ we can find an $F$-disjoint path $z \leadsto w$. By Proposition~\ref{prp:outerplanar}, this implies that $d(\hz,w) \leq 1$. Then (1): $d(\hz, y) \leq 1 + d(w, y) \leq 1 + d(s, y) = 1 + d(v, y) - 1 = d(v, y)$. Applying Proposition~\ref{prp:cycle:shortest} to $z$, $x$, and $C$, and to $z$, $x$, and $F$, we observe that (2): $d(\hz, x) \geq d(v, x)$. Since $d(v,x) = d(v,y)$ by definition, the equality $d(\hz,x) = d(\hz,y)$ holds only if equality holds in both (1) and (2). Then $w = s$ and $\hz = v$. 
Moreover, as $v$ is the representative of $z$ on $C$, $d(z,v) < d(z,s)$ and $d(\hz,v) < d(\hz,s)$. This, together with the assumption that no shortest path $z \leadsto y$ intersects $F$ and with the definition of $F$, implies the existence of a subdivision of $K_{2,3}$ as a subgraph, a contradiction.

Suppose then that some shortest path $z \leadsto y$ intersects $F$. By Proposition~\ref{prp:cycle:shortest}, there is a shortest path $z \leadsto y$ that contains $\hz$ or, if $\hz$ is a midpoint on an edge, contains one of the endpoints of this edge. By Proposition~\ref{prp:cycle:shortest} and the earlier observation that any path $z \leadsto x$ intersects $F$, the same holds for a shortest path $z \leadsto x$. Then $0 > d(z, y) - d(z, x) = d(\hz, y) - d(\hz, x)$. Hence, $\hz$ resolves $x$ and $y$. This proves the intermediate result.

We now claim that $L_{1}$ has more than one representative on $F$. Let $\hz_{1}$ be the representative of $z_{1}$ on $F$. Note that by the definition of an implied cycle, $z_{1}$ has a shortest path to $x$ and to $y$ that contains $v$. Since $v \in F$, there is a shortest path from $z_{1}$ to $x$ and from $z_{1}$ to $y$ that intersects $F$. Hence, it follows from Proposition~\ref{prp:cycle:shortest} that if $\hz_{1} \in V(G)$, then there are shortest paths from $z_{1}$ to $x$ and to $y$ that both contain $\hz_{1}$; if $\hz_{1}$ is a midpoint of an edge of $F$, then there are such shortest paths that both contain an endpoint of that edge. As $z_{1}$ does not resolve $x$ and $y$, neither does $\hz_{1}$. Since $\hz$ does resolve $x$ and $y$, $\hz \not= \hz_{1}$, and thus $L_{1}$ has more than one representative on $F$.

To prove that $F$ is an extreme face of $L_1$ and $C'$, we note that there is a path in the weak dual between $C'$ and $F$ that contains only faces that are topologically contained in $C$. Hence, it would suffice to prove that no face $D$ that is topologically contained in $C$ has more than one representative of $L_{1}$ on it. Observe that $L$ has at least two representatives on $D$ by Lemma~\ref{lem:cycle:two}, since $L$ is a resolving set and thus satisfies Requirement~\ref{req:g}. Moreover, $L$ has at most two representatives on this face by Corollary~\ref{cor:three-rep}, as the face is topologically contained in $C$ and $L$ has two representatives on $C$. Hence, $L$ has two representatives on any face that is topologically contained in $C$; in particular, $L_1$ and $L_2$ each have one representative on each such face. Therefore, $F$ is an extreme face.

Finally, we need to show that the extreme representative $e$ of $L_{1}$ and $C'$ that lies on $F$ resolves $x$ and $y$. Since $e$ is on $F$, it has $C$-disjoint paths to both $v$ and $s$. Hence, $e$ cannot have a $C$-disjoint path to $x$, as such a path could be extended to a $C$-disjoint path to $u$, contradicting Proposition~\ref{prp:outerplanar}. Suppose that there is a shortest path from $e$ to $x$ that contains $s$. Then, by Proposition~\ref{prp:cycle:w}, $d(e,x) = d(e,s) + d(s,x) > d(e,s) + d(s,y) \geq d(e,y)$, and thus $x$ and $y$ are resolved. Suppose then that there is a shortest path from $e$ to $x$ that contains $v$. Recall that $\hz$ resolves $x$ and $y$, and in particular that $d(\hz, y) < d(\hz, x)$. Moreover, as the representative of $z$ on $C$ is $v$ and any path from $z$ to $C$ contains a vertex of $F$, the representative of $\hz$ on $C$ is $v$ as well. 
Note that the definition of extreme representative implies that $e$ is furthest away from $v$ among all representatives of $L_{1}$ on $F$. Therefore, the shortest path from $e$ to $x$ that contains $v$ also contains $\hz$, and thus:
$$d(e,x) = d(e,\hz) + d(\hz,v) + d(v, x) = d(e, \hz) + d(\hz,x) > d(e,\hz) + d(\hz,y) \geq d(e,y),$$
Hence, $e$ resolves $x$ and $y$.
\qed\end{proof}

\subsection{A Characterization for Outerplanar Graphs} \label{sec:sub:charac}

Using the notions and results of the previous sections, we can generalize Theorem~\ref{thm:tree} to outerplanar graphs. This is a crucial result, since it characterizes resolving sets in a manner that allows for the use of dynamic programming.

Let $G$ be a connected outerplane graph. We will show that if $L \subseteq V(G)$ is a resolving set, then it satisfies the following requirement:

\begin{Req} \label{req:cycle}
Let $z_1, z_2 \in L$ and $x, y \in V(G)$ be four distinct vertices that imply a cycle $C$ of $G$ such that $L$ has exactly two representatives on $C$, and let $C'$ be a face that is topologically contained in $C$. Then one of the extreme representatives of $L$ and $C'$ resolves $x$ and $y$.
\end{Req}

For example, Figure~\ref{fig:extremeres} satisfies the requirement with respect to $L$ and $C=C'$, since the extreme representative $e$ of $L$ and $C=C'$ resolves $x$ and $y$.

This requirement leads to one of the central results of this paper.

\begin{theorem}\label{thm:main}
Let $G$ be a connected outerplane graph with at least three vertices. Then a set $L \subseteq V(G)$ is a resolving set of $G$ if and only if it satisfies Requirement~\ref{req:g} and~\ref{req:cycle}.
\end{theorem}
\begin{proof}
To see that Requirement~\ref{req:g} and~\ref{req:cycle} are necessary, suppose that $L$ is a resolving set. It was argued before the statement of Requirement~\ref{req:g} that $L$ satisfies this requirement. The fact that $L$ satisfies Requirement~\ref{req:cycle} is immediate from Lemma~\ref{lem:cycle:neighbors}.

We now show that Requirement~\ref{req:g} and~\ref{req:cycle} are sufficient. Suppose that $L \subseteq V(G)$ satisfies Requirement~\ref{req:g} and~\ref{req:cycle}, and choose any $x, y \in V(G)$. We show that there exists a $z \in L$ that resolves the pair $x,y$. Suppose for sake of contradiction that $L$ does not resolve $x,y$. 

Using the same arguments as in the proof of Theorem~\ref{thm:tree}, $L$ is non-empty. Choose $z_1 \in L$ arbitrarily. As in Theorem~\ref{thm:tree}, let $v$ be the bifurcation point of $z_1,x,y$, and let $v_1, v_2$ be successors of $v$ on some shortest paths $v \leadsto x, v \leadsto y$ respectively. By Requirement~\ref{req:g}, there is a $z_2 \in L$ such that, without loss of generality, $d(z_2, v_1) \leq d(z_2, v)$.  By assumption, neither $z_{1}$ nor $z_{2}$ resolves $x,y$.

Suppose that a shortest path $z_1 \leadsto x$ intersects a shortest path $z_2 \leadsto y$ on a vertex $w$. Let $w'$ be the bifurcation point of $w,x,y$. Note that there exists a shortest path $z_1 \leadsto x$ that intersects a shortest path $z_2 \leadsto y$ on $w'$. Using similar arguments as in Lemma~\ref{lem:cycle}, it follows that $w'$ is also on a shortest path $z_1 \leadsto y$ and a shortest path $z_2 \leadsto x$. But then $w'$ is the bifurcation point of $z_1, x, y$, \ie $w' = v$. Moreover, since $v$ is on a shortest path $z_{2} \leadsto x$, $d(z_2, v_1) > d(z_2, v)$, contradicting the choice of $z_2$. Hence, no two shortest paths $z_1 \leadsto x$, $z_2 \leadsto y$ intersect.

It follows from Lemma~\ref{lem:cycle} that the vertices $z_1, z_2, x, y$ imply a cycle $C$. Let $v, t, s, u$ be the defining representatives of $C$. Let $C'$ be any face that is topologically contained in $C$. By Lemma~\ref{lem:cycle:two}, $L$ has at least two representatives on $C'$, because $L$ satisfies Requirement~\ref{req:g}. Then $L$ has at least two representatives on $C$ by Corollary~\ref{cor:three-rep}.

If $L$ has at least three representatives on $C$, then Lemma~\ref{lem:implied:three} shows that $L$ resolves $x$ and $y$ and the theorem follows. Hence, from now on, we assume that $L$ has exactly two representatives on $C$. Then from the statement of Requirement~\ref{req:cycle} there is an extreme representative $\hz$ of $L$ and $C'$ that resolves $x$ and $y$. Let $z \in L$ be the corresponding vertex. It remains to show that $z$ resolves $x$ and $y$ as well.

Let $L_{1} \subset L$ be the set of vertices that have representative $v$ on $C$ (cf.~Definition~\ref{def:extremerepr}). Without loss of generality, $z \in L_{1}$ and $d(\hz, y) < d(\hz,x)$. Note that $\hz$ does not lie on $C$: by assumption, $L$ has two representatives on $C$, neither of which resolves $x$ and $y$.

We now show that $L_{1}$ and $C'$ has an extreme face. Suppose for sake of contradiction that $L_{1}$ has only one representative on the biconnected component that contains $C'$. Then this representative is the extreme representative of $L_{1}$ and $C'$, and thus it is $\hz$. Since any path from $z$ to $x$ and $y$ contains $\hz$, $d(z, x) = d(z, \hz) + d(\hz, x)$ and $d(z, y) = d(z, \hz) + d(\hz, y)$. However, $z$ does not resolve $x$ and $y$ by assumption, whereas $\hz$ does, a contradiction. Therefore, $L_{1}$ has more than one representative on the biconnected component, and there exists an extreme face of $L_{1}$ and $C$. Denote the extreme face by $F$.

We need several auxiliary results.

\medskip\noindent
{\bf Claim~1:} Any path from $z$ to $C$ contains a vertex of $F$.\\
{\bf Proof:}
Let $F'$ be the face adjacent to $F$ that is between $F$ and $C'$ in the weak dual (note that possibly $F'=C'$). By definition, some vertex $z' \in L_{1}$ has a representative $\hz'$ on $F$ that is distinct from $\hz$. Suppose that $\hz$ is on the edge separating $F$ and $F'$ (the \emph{boundary}). Then $\hz'$ is also on the boundary, as by the definition of an extreme representative it must be at least as close to the common representative on $C'$ as $\hz$ is. Observe that $z$ and $z'$ have the same representative on $F'$. If this common representative does not lie on the boundary between $F$ and $F'$, then Proposition~\ref{prp:outerplanar} and Proposition~\ref{prp:cycle:shortest} imply that any shortest path from $z$ or $z'$ to $F$ intersects $F'$. Hence, Proposition~\ref{prp:cycle:shortest} implies that $z$ and $z'$ have the same representative on $F$, a contradiction. Therefore, the common representative does lie on the boundary. But then this common representative equals $\hz$ and $\hz'$. Thus, $\hz=\hz'$, a contradiction. Hence, $\hz$ is not on the boundary between $F$ and $F'$. Suppose now that there is a path from $z$ to $C$ that contains no vertex of $F$. Then it could be extended to $F$-disjoint paths from $z$ to both vertices on the boundary between $F$ and $F'$. Since $\hz$ is not on this boundary, we obtain a contradiction to Proposition~\ref{prp:outerplanar}. The claim follows.\qquad \#

\medskip\noindent
{\bf Claim~2:} There is a $C$-disjoint path from $z$ to $s$.\\
{\bf Proof:} Observe that Claim~1 together with Proposition~\ref{prp:cycle:shortest} implies that the representative of $\hz$ on $C$ is $v$. Since $d(\hz, y) < d(\hz, x)$, the reasoning in the second paragraph of Lemma~\ref{lem:cycle:neighbors} implies that there is a $C$-disjoint path from $\hz$ to $y$. There is also a $C$-disjoint path from $z$ to $\hz$, and combining them yields a $C$-disjoint path from $z$ to $y$. Using a shortest path from $s$ to $y$, this path can be further extended to a $C$-disjoint path from $z$ to $s$.\qquad\#

\medskip\noindent
{\bf Claim~3:} Any path from $z$ to $x$ must contain a vertex of $F$.\\
{\bf Proof:} Suppose not, and consider a path from $z$ to $x$ that does not contain a vertex of $F$. If this path is $C$-disjoint, then it can be extended to a $C$-disjoint path from $z$ to $u$. This path, together with the $C$-disjoint path from $z$ to $s$ which exists by Claim~2, contradicts Proposition~\ref{prp:outerplanar}. Therefore, the path is not $C$-disjoint. Then there is a path from $z$ to $C$ that contains no vertex of $F$, a contradiction to Claim~1. So any path from $z$ to $x$ contains a vertex of $F$.\qquad\#

\medskip
We are now ready to prove the theorem. If $\hz$ is a regular vertex, then applying Claim~3 and Proposition~\ref{prp:cycle:shortest} implies that any shortest path from $z$ to $x$ contains $\hz$, so $d(z, y) \leq d(z, \hz) + d(\hz, y) < d(z, \hz) + d(\hz, x) = d(z, x)$, and thus $z$ resolves $x$ and $y$.
Similarly, if $\hz$ is a midpoint, then $d(z, y) \leq d(z, \hz) + d(\hz, y) - 1 < d(z, \hz) + d(\hz, x) - 1 = d(z, x)$, and thus $z$ resolves $x$ and $y$. The theorem follows.
\qed\end{proof}

\section{Algorithm for Outerplanar Graphs} \label{sec:algorithm}
In this section, we prove that {\metricdp} can be solved in polynomial time on outerplane graphs, and thus also on outerplanar graphs. 
We consider first the data structures that support the algorithm. The algorithm will build up a resolving set in a dynamic-programming fashion. Therefore, we need to find a suitable order in which to process the outerplane graph. This order will be given by a (rooted) annotated generalized dual tree, which is closely related to the planar dual of the outerplane graph. The tree has in its vertex set all cut and pendant vertices of the outerplane graph and all face vertices of the dual. The precise structure and the way it is annotated is considered in more detail in Section~\ref{sec:algorithm:dual}, together with its most crucial properties.

When traversing the tree, we need to combine the information of the children of a tree vertex and send this combination to the parent of that vertex. Moreover, the tree vertex may expect certain vertices of the partial resolving set to be present in the yet unprocessed part of the tree. These `requests' must also be sent to the parent vertex. The combined information of descendants and requests to the parents are made through so-called boundary conditions, which are defined in Section~\ref{sec:algorithm:boundary}. We also need to properly combine the information of the children of a vertex and to satisfy Requirement~\ref{req:g} and~\ref{req:cycle}. This is mainly accomplished by configurations, defined in Section~\ref{sec:algorithm:configurations}. Several properties of configurations are also proved there.

Finally, we give the algorithm in Section~\ref{sec:algorithm:algorithm}. The correctness of the algorithm follows from the characterization of resolving sets in outerplane graphs of Section~\ref{sec:algorithm:charac}.

\subsection{Generalized Dual Tree} \label{sec:algorithm:dual}
We describe the order in which the outerplane graph will be processed by the algorithm. For this we use the generalized dual tree. 

\begin{figure}
\scalebox{0.47}{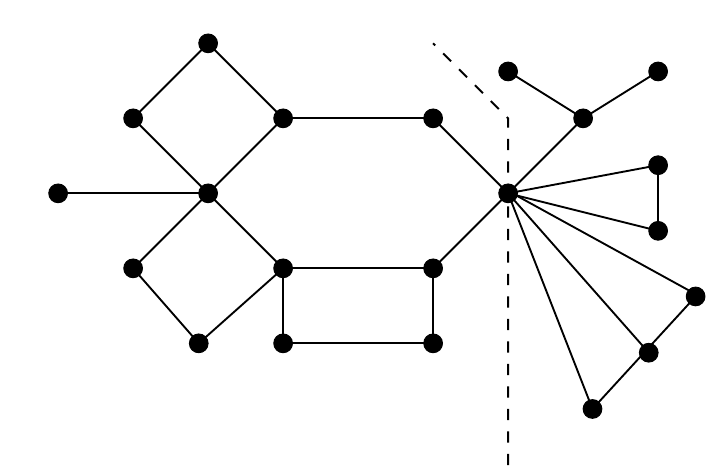 } \hspace{0.3cm}
\scalebox{0.47}{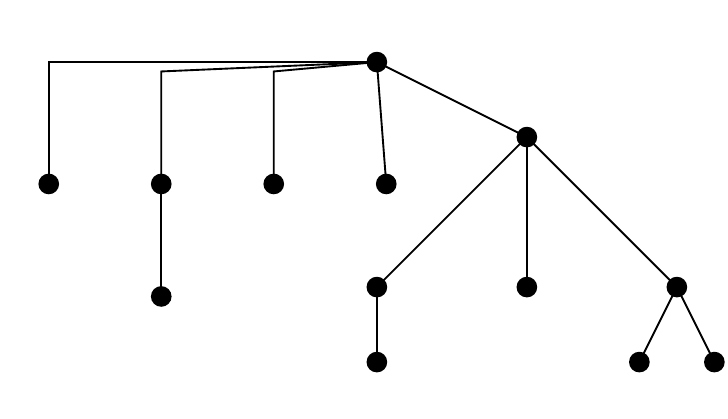 }
\caption{An outerplane graph $G$ and its generalized dual tree $T$. Note that for $v_1$, the faces $f_{v_1,C}$ are $F_1$, $F_5$, and $F_6$ respectively. 
The dashed line indicates how $G$ is divided to $B(F_1, v_1)$ and
$B(v_1, F_1)$. The sets on the edges of $T$ indicate the values of $s(e')$.
\label{fig:extdualtree}}
\end{figure}

\begin{definition}
Let $G$ be a connected outerplane graph. For each cut vertex~$c$ and each nontrivial biconnected component $C$ of $G$ that contains $c$, let $f_{c,C}$ be an arbitrary face of $C$ that contains the cut vertex. 
Then a \emph{generalized dual tree} $T=(V',E')$ of $G$ is defined as follows.
$V'$ is the union of the set of faces of $G$ (except the outer face), the set of cut vertices of $G$, and the set of vertices of $G$ of degree~$1$.
There is an edge in $E'$ between
\begin{titemize}
\item two vertices corresponding to two faces if the faces share an edge of $G$;
\item two cut vertices if these vertices are adjacent in $G$;
\item two vertices of degree~$1$ if these vertices are adjacent in $G$;
\item a cut vertex and a vertex of degree~$1$ if these vertices are adjacent in $G$;
\item a cut vertex $c$ contained in a nontrivial biconnected component $C$ and the vertex corresponding to $f_{c,C}$.
\end{titemize}
Let an arbitrary vertex of $T$ be the root, denoted by $v'_r$.
\end{definition}
Observe that a fixed outerplane graph might have many generalized dual trees, depending on the choices made for the faces $f_{c,C}$. For the purposes of this paper, the precise faces $f_{c,C}$ chosen are immaterial. Therefore, without loss of generality, we will speak of \emph{the} generalized dual tree $T$ of $G$. Figure~\ref{fig:extdualtree} illustrates the definition.

Note that the generalized dual tree of an outerplanar graph $G$ contains the weak dual as an induced subgraph, because the weak dual has the faces of $G$ (except the outer face) as vertices, and two faces are adjacent if they share an edge of $G$. It then follows that the generalized dual tree is indeed a tree by construction, because the weak dual of $G$ is a forest. Moreover, according to the definition of a generalized dual tree, a cut vertex is a vertex of both $G$ and~$T$.

We now annotate the generalized dual tree. We associate a subset of $V(G)$ with each vertex and each edge of $T$. If $v' \in V(T)$ is a face, then the set $s(v')$ consists of the vertices on the face. If $v'$ is a cut vertex or a pendant vertex, then $s(v')$ consists of that vertex. Let $e'=(v',p')$ be an edge of $T$, where $p'$ is the parent of $v'$. If $s(v')$ and $s(p')$ correspond to a cut or pendant vertex of $G$, then $s(e') = s(p')$. Otherwise, at least one of $s(v'),s(p')$ is a face, and we set $s(e') = s(v') \cap s(p')$.

When traversing the annotated generalized dual tree, it will be useful sometimes to combine the sets $s$ for all descendants or ancestors of a vertex of the tree. To this end, we define the function $B$. Removing an edge $\edge{v'}{w'}$ divides $T$ into two components,
$T_{v'}$ and $T_{w'}$, where $T_{v'}$ is the one containing $v'$.
Define $B(v', w')$ as the subgraph of $G$ corresponding to $T_{v'}$.
Formally, it is the 
subgraph of $G$ induced by $\bigcup_{u' \in V(T_{v'})} s(u')$.
Note that $B(v',w')$ and $B(w',v')$ are two different subgraphs of $G$ (see Figure~\ref{fig:extdualtree}). Moreover, if $v', w'$ are adjacent faces, then the subgraphs share two vertices and an inner edge; if $v'$
is a face and $w'$ a cut vertex (or the other way around), then the
subgraphs share one vertex; otherwise, they do not intersect. To avoid this possibly nonempty intersection, define $B^-(v', w')$ as the subgraph of $G$ induced by $V(G) \setminus
V(B(w', v'))$. Then we can divide $G$ into two nonintersecting subgraphs,
$B^-(v', w')$ and $B(w', v')$.

Finally, the notion of representative can be extended to the generalized dual tree. We can define the representative of $z \in V(G)$ on $v' \in V(T)$ or on $e' \in E(T)$ as the representative of $z$ on $s(v')$ or on $s(e')$, respectively.

We now state three important, straightforward lemmas about the generalized dual tree and its annotation. The first lemma is immediate from the definitions.
\begin{lemma} \label{lem:dualstructure}
Let $e' = \edge{v'}{w'} \in E(T)$. Then $B(v', w')$ and $B(w', v')$ are connected subgraphs of $G$, and any path from $B(v', w')$ to $B(w', v')$ intersects $s(e')$.
\end{lemma}

\noindent
The second lemma is also immediate from the definitions.

\begin{lemma} \label{lem:three-rep:separator}
Let $e' = \edge{v'}{w'} \in E(T)$, let $L$ be a set of vertices, and let $Z$ be the set of representatives of $L \cap B(v',w')$ on $e'$. Then
\begin{itemize}
\item $g(v,L) \cap V(B(w',v')) = g(v,Z \cup (L \cap V(B(w',v'))))  \cap V(B(w',v'))$ for any $v \in B(w',v')$;
\item $L$ resolves $x,y \in V(B(w',v'))$ if and only if $Z \cup (L \cap V(B(w',v')))$ does.
\end{itemize}
\end{lemma}

\noindent
In the third lemma, we prove that if both endpoints and the midpoint of an inner edge are representatives of a set of vertices, then we do not need to remember that the midpoint is a representative.

\begin{lemma} \label{lem:midpoint}
Let $e' = \edge{v'}{w'} \in E(T)$ correspond to an inner edge $e = \edge{v_{1}}{v_{2}}$ with midpoint $v_{e}$, let $L \subseteq V(G)$, and let $\{v_{1},v_{2},v_{e}\}$ be the set of representatives of $L \cap V(B(v',w'))$ on $e'$. Then:
\begin{itemize}
\item $g(v,L)  \cap V(B(w',v')) = g(v,\{v_{1},v_{2}\} \cup (L \cap V(B(w',v'))))  \cap V(B(w',v'))$ for any $v \in B(w',v')$;
\item $L$ resolves $x,y \in V(B(w',v'))$ if and only if $\{v_{1},v_{2}\} \cup (L \cap V(B(w',v')))$ does.
\end{itemize}
\end{lemma}
\begin{proof}
The first part follows from the fact that a shortest path from a vertex in $L \cap V(B(v', w'))$ to a vertex $v$ in $B(w', v')$ contains either $v_1$ or $v_2$ by Lemma~\ref{lem:dualstructure}.

Consider the second part and suppose that $x,y$ are resolved by $v_{e}$, but not by $v_{1}$. Then $d(v_1, x) = d(v_1, y)$ and \wloge $d(v_e, x) < d(v_e, y)$. Then
$$ {\textstyle \frac{1}{2}}+\min\{d(v_1, x), d(v_2, x)\} < {\textstyle \frac{1}{2}}+\min\{d(v_1, y), d(v_2, y)\}. $$
Putting these equations together implies that $d(v_2, x) < d(v_2, y)$, that is, $x,y$ are resolved by $v_{2}$.
\qed\end{proof}
Following this lemma, if $L \cap V(B(v',w'))$ has representatives $\{v_1,v_2,v_e\}$ on $e'$, where $e'$ corresponds to an inner edge $e$ and $v_{e}$ is the midpoint of $e$, then we can equivalently say that it has representatives $\{v_1,v_2\}$, without losing any expressive power. That is, our ability to determine the function $g$ restricted to $B(w',v')$ is unaffected, as well as our ability to determine which vertices are resolved in $B(w',v')$. We use this equivalence extensively later on and essentially treat $\{v_1,v_2,v_e\}$ as being equal to $\{v_1,v_2\}$.

\subsection{Boundary Conditions} \label{sec:algorithm:boundary}
We now define the `states' or `indices' of the dynamic-programming table in our algorithm. We call these `states' the \emph{boundary  conditions}. Throughout this section, let $G$ be a connected outerplane graph, let $T$ be the generalized dual tree of $G$ rooted at an arbitrary vertex $v'_{r}$, and let $s$ be the annotation of $T$ as defined before.

Let $v' \in V(T)$, let $p'$ be its parent, and let $e' = \edge{v'}{p'} \in E(T)$. Define the set $X_{e'}$ as follows: if $e'$ corresponds to an inner edge $e$ of $G$, then let $X_{e'}$ be the union of $s(e')$ and $\{v_{e}\}$, where $v_{e}$ is the midpoint of $e$; otherwise, let $X_{e'} = s(e')$. Then a boundary condition can be defined as follows.

\begin{definition}
Let $v' \in V(T)$, let $p'$ be its parent, and let $e' = \edge{v'}{p'} \in E(T)$.
A \emph{boundary condition} ${\bf t}$ of $e'$ is a tuple consisting of the following elements: 
\begin{itemize}
\item $t_b, t_{rl}, t_{ru} \subseteq X_{e'}$,
\item $t_{el}, t_{eu} \in \mathcal{P}^2(V(G)) \cup \{ 0\}$, and 
\item $t_{v_1}, t_{v_2} \in \{0, 1\}$,
\end{itemize}
where $t_{v_2}$ is included only if $|X_{e'}| > 1$.
\end{definition}

Given this definition, we need to define what it means for a (partial) solution to satisfy the boundary condition.

\begin{definition} \label{def:boundary:adheres}
Let $v' \in V(T)$, let $p'$ be its parent, and let $e' = \edge{v'}{p'} \in E(T)$.
We say that a set $L \subseteq V(G)$ of vertices \emph{adheres} to a boundary condition ${\bf t}$ of $e'$ if:
\begin{itemize}
\item $t_{b}$ is equal to the set $L \cap s(e')$ if at least one of $v',p'$ is a face, and $t_{b}$ is $\emptyset$ otherwise.
\item $t_{rl}$ is equal to the set of representatives of $L \cap V(B(v', p'))$ on $s(e')$. (Note that $t_{b} \subseteq t_{rl}$.)
\item $t_{el}$ is equal to $\emptyset$ if at least one of $v',p'$ is not a face. Otherwise, $t_{el}$ is equal to:
\begin{itemize}
\item $0$ if there is a set $L' \subseteq L \cap V(B(v', p'))$ of size two such that $L'$ has two representatives on $e'$ and shortest paths from $L'$ to $e'$ intersect a face $y'$ on which $L$ has at least three representatives;
\item the representative on $v'$ furthest from the vertex in $t_{rl}$ if $L$ has at least three representatives on $v'$ and $|t_{rl}| = 1$;
\item the extreme representatives of $L$ and $p'$ that are in $B(v', p')$ otherwise.
\end{itemize}
\item $t_{ru}$ and $t_{eu}$ are as $t_{rl}$ and $t_{el}$ respectively, but with $v'$ and $p'$ interchanged.
\item $t_{v_i} = |g(v_i, L) \cap V(B(v',p'))|$ for $i=1$ if $|X|=1$ and for $i=1,2$ otherwise. (Note that this may depend on both $L \cap V(B^-(v', p'))$ and $L \cap V(B(p', v'))$.)
\end{itemize}
\end{definition}
For $t_{rl}$ and $t_{ru}$ we treat $\{v_1, v_2\}$ as being equal to $\{v_1, v_2, v_e\}$, as explained below Lemma~\ref{lem:midpoint}. Hence, if $e'$ corresponds to an inner edge, then $t_b, t_{rl}, t_{ru} \subset X_{e'}$.

The purpose of $t_{el}$ is to contain the extreme representatives that the algorithm needs to verify Requirement~\ref{req:cycle}. However, we can not simply say that $t_{el}$ must be equal to the set of extreme representatives, as the algorithm does not always know them. Therefore, we need the more complicated definition given above. The following lemmas show that either $t_{el}$ contains the extreme representatives, or Corollary~\ref{cor:three-rep} indicates that Requirement~\ref{req:cycle} is satisfied.

\begin{proposition}\label{prp:tel:three}
Let $v',p'$ correspond to faces that share an edge, let ${\bf t}$ be a boundary condition of $\edge{v'}{p'}$ with $t_{el} = 0$, and let $L \subseteq V(G)$ adhere to ${\bf t}$. Then any implied cycle of $L$ that contains $p'$ or $v'$ satisfies Requirement~\ref{req:cycle}.
\end{proposition}
\begin{proof}
Let $C$ be an implied cycle of $L$ that contains $p'$ or $v'$. Note that if $L$ has at least three representatives on $C$, then Requirement~\ref{req:cycle} is trivially satisfied. Combined with the definition of an implied cycle, we may assume that $L$ has exactly two representatives on $C$. 
Since $L$ adheres to ${\bf t}$ and $t_{el}=0$, by definition there is a face $y'$ on which $L$ has at least three representatives.
If $y'$ is inside $C$, then it follows from Corollary~\ref{cor:three-rep} that $L$ has at least three representatives on $C$, a contradiction. Hence, $y'$ is not inside $C$.

Face $y'$ and cycle $C$ are a vertex and a connected component of the generalized dual tree, so there is a unique path between them. Let $\edge{w'}{u'}$ be the last edge of this path so that $u'$ is inside $C$ and $w'$ is not (note that possibly $\edge{w'}{u'} = \edge{v'}{p'}$). Let $L'$ be as in the definition of the case $t_{el} = 0$. Then $L'$ must have at least two representatives on $\edge{w'}{u'}$---otherwise, $L'$ would not have two representatives on $\edge{v'}{p'}$ (which corresponds to an edge or a chord of $C$). Observe, however, that not all representatives of $L$ on $C$ are on the edge of $C$ corresponding to $\edge{w'}{u'}$. By Proposition~\ref{prp:cycle:repr} the representatives of $L$ on $C$ are not adjacent, but have on a path between them one of the other two defining representatives of the implied cycle $C$.
Hence, $L$ has at least three representatives on $C$ (two on $\edge{w'}{u'}$, and a vertex among $v,t$ that is not on the edge of $C$ corresponding to $\edge{w'}{u'}$), a contradiction. The proposition follows.
\qed\end{proof}

\begin{proposition}\label{prp:tel:ext}
Let $v',p'$ correspond to faces that share an edge, let ${\bf t}$ be a boundary condition of $\edge{v'}{p'}$ with $t_{el} \neq 0$, and let $L \subseteq V(G)$ be a set of vertices that adheres to ${\bf t}$ and that has less than three representatives on $p'$.
Then $t_{el}$ is equal to the set of the extreme representatives of $L$ and $p'$ that are in $B(v', p')$.
\end{proposition}
\begin{proof}
The definition of $t_{el}$ contains three cases when $v'$ and $p'$ are faces. We have excluded the first case by assumption, and in the third case the proposition follows by definition. Consider the second case,
i.e.~suppose that $t_{el} \neq 0$, that $L$ has at least three representatives on $v'$, and that $|t_{rl}| = 1$.
It suffices to show that $v'$ is the relevant extreme face, i.e.~that $L$ contains two vertices that have different representatives on $v'$ but the same representative on $p'$. Then it follows by definition that $t_{el}$ contains the extreme representative.

Let $L_s = \{z_1, z_2, z_3\} \subseteq L$ have different representatives on $v'$ such that $t_{el}$ contains the representative of $z_1$ on $v'$. Then $z_1 \in V(B(v', p'))$. Since $L$ (and thus also $L_{s}$) has less than three representatives on $p'$ by assumption, some of $z_1, z_2, z_3$ have the same representative on $p'$. If $z_2$ or $z_3$ has the same representative on $p'$ as $z_1$ does, then the proposition follows. Otherwise, $z_2$ and $z_3$ have the same representative on $p'$, which is different from the representative of $z_{1}$ on $p$.
Since $|t_{rl}| = 1$, both $z_2$ and $z_3$ are in $V(B^{-}(p', v'))$. This contradicts that $z_2$ and $z_3$ have different representatives on $v'$.
\qed\end{proof}

The following is needed to show that the algorithm runs in polynomial time.

\begin{lemma} \label{lem:boundary}
For any edge $e'$ of $T$, there are $O(n^{4})$ boundary conditions. Moreover, all boundary conditions  of $e'$ can be enumerated in polynomial time. Finally, any resolving set adheres to some boundary condition on $e'$.
\end{lemma}
\begin{proof}
Since $|X_{e'}| \leq 3$, there are at most $2^3$ choices for each of $t_b$, $t_{rl}$, and $t_{ru}$. For both $t_{eu}$ and $t_{el}$ there are $O(n^2)$ choices, and for both $t_{v_1}$, $t_{v_2}$ there are two choices. Multiplying gives that there are $O(n^4)$ boundary conditions. They can be enumerated in the same time. The final statement of the lemma follows immediately from Definition~\ref{def:boundary:adheres}.
\qed\end{proof}

\subsection{Configurations} \label{sec:algorithm:configurations}
The boundary conditions defined in the previous section define the states of the dynamic-programming table. The main problem we are faced with now is to compute the table entry for the boundary condition of an edge $\edge{v'}{p'}$ using the entries for the boundary conditions of the edges $\edge{w'}{v'}$ of the children $w'$ of $v'$. To this end, we define a new structure, called a \emph{configuration}, that determines what the solution for $s(v')$ looks like for each vertex $v' \in V(T)$. The definition of a configuration depends on whether $v'$ corresponds to a face (and what type of face), or to a cut or pendant vertex of $G$ (e.g.\ $F_1$, $v_{16}$ or $v_{17}$ in Figure~\ref{fig:extdualtree}, respectively). 
When $v'$ is a face, the definition splits into two different cases. Recall that by Lemma~\ref{lem:cycle:two}, each face will have at least two representatives. Therefore, we distinguish two types of configurations for faces, depending on whether a face has two representatives or more than two. This leads to a total of three configuration types (two when $v'$ is a face, and one otherwise).

As before, throughout this section, let $G$ be a connected outerplane graph, let $T$ be the generalized dual tree of $G$ rooted at an arbitrary vertex $v'_{r}$, and let $s$ be the annotation of $T$ as defined before. Moreover, let $v'$ be an arbitrary vertex of $T$.

We first introduce the following notion for the case when $v'$ corresponds to a face.

\begin{definition} \label{def:extended}
Let $z \in V(G)$ and let $\hz$ be the representative of $z$ on $v'$, where $v'$ is a face of $G$. Then the \emph{extended representative} of $z$ on $v'$ is the pair $(\hz,w')$, where $w'$ is $v'$ if $z \in s(v')$ and $w'$ is the neighbor of $v'$ in $T$ such that $z \in V(B^{-}(w',v'))$ otherwise.
\end{definition}

Observe that for a fixed representative $\hz$, there are at most four values of $w'$ such that $(\hz, w')$ is a valid extended representative, namely $v'$, the at most two faces that contain $\hz$ and share an edge with $s(v')$, and (if $\hz$ is a cut vertex) the vertex of $T$ corresponding to this cut vertex.

In the following subsections, we give the definition of each configuration type.

\subsubsection{Configurations of type I}\label{sec:algorithm:configurationsI}
First, we define a configuration for the case that $s(v')$ is a cut or pendant vertex $v$. We note that we do not need to consider cut or pendant vertices when satisfying Requirement~\ref{req:cycle}, as this Requirement concerns only cycles and faces. We only need to verify Requirement~\ref{req:g}. Also, we need to know whether $v$ is a landmark or not. 

\begin{definition}
Let $v' \in V(T)$ correspond to a cut or pendant vertex $v$.
A \emph{configuration of type I} on $v'$ consists of a boolean variable indicating whether $v$ is a landmark, and at most one vertex that is in $g(v,\cdot)$. A set $L \subseteq V(G)$ \emph{adheres} to a configuration of type I if $v \in L$ if and only if the configuration specifies this, and $g(v,L) = \emptyset$ if the configuration specifies this or $g(v,L)$ consists of the single vertex specified by the configuration.
\end{definition}

\begin{lemma} \label{lem:conf:I}
Let $v' \in V(T)$ correspond to a cut or pendant vertex $v$, let $\mc{C}$ be a configuration of type I on $v'$, and let $L \subseteq V(G)$ adhere to $\mc{C}$.
Then $\mc{C}$ determines the boundary conditions adhered to by $L$ for all edges $\edge{v'}{w'}$, where $w'$ is a neighbor of $v'$ in $T$, and these boundary conditions can be computed in polynomial time.
Finally, all configurations of type I can be enumerated in polynomial time.
\end{lemma}
\begin{proof}
Suppose that $w'$ is a child of $v'$ (similar arguments apply when $w'$ is the parent of $v'$), and consider the boundary condition ${\bf t}$ of $e' = \edge{w'}{v'}$. Note that $s(e') = s(v') = \{v\}$. If $w'$ is a face and $\mc{C}$ specifies that $v$ is a landmark, then $t_b = \{v\}$; otherwise $t_b = \emptyset$. Because $v'$ is not a face, $t_{el} = t_{eu} = \emptyset$. Since $g(v, L)$ is known, it is easy to determine $t_{v_1}$.

The component $t_{rl}$ of ${\bf t}$ is $\{v\}$ if $v$ is a landmark and $v \in B(w', v')$, or if $|g(v,L) \cap V(B(w',v'))| < |N(v) \cap V(B(w',v'))|$; otherwise, $|g(v,L) \cap V(B(w',v'))| = |N(v) \cap V(B(w',v'))| = 1$ and $t_{rl}$ is $\emptyset$. To see this, note that any vertex in $L \cap B(w', v')$ has representative $v$ on $v'$, so the question is whether there is a vertex in $L \cap B(w', v')$. A vertex $w \in V(B(w', v'))$ is not in $g(v, L)$ if and only if there is a $z \in L$ such that $d(z, w) \leq d(z, v)$ and thus $z \in V(B(w', v'))$. A similar argument holds for $t_{ru}$; however, if the configuration indicates that $v$ is a landmark, then $t_{ru} = \{v\}$ regardless of $g(v, L)$.

Finally, there are at most $O(n)$ configurations of type I, and enumerating all of these can be done in $O(n)$ time.
\qed\end{proof}

\subsubsection{Configurations of type II} \label{subp:II}
Suppose that there are two representatives on $v'$. Then the algorithm has to verify Requirement~\ref{req:cycle}. We immediately run into two practical issues. First, the algorithm does not know all landmarks yet, so how do we detect implied cycles that could violate Requirement~\ref{req:cycle}? Second, the boundary condition on $\edge{w'}{v'}$ for some neighbor $w'$ of $v'$ in $T$ does not specify the extreme landmarks in all cases. Can we verify Requirement~\ref{req:cycle} then? We address each of these issues in turn.

We start by detecting implied cycles. For this, we can use extreme representatives.

\begin{lemma} \label{lem:identify}
Let $L \subseteq V(G)$, let $z_{1},z_{2} \in L$ and $x,y \in V(G)$ be four distinct vertices that imply a cycle $C$ of $G$ such that $L$ has exactly two representatives on $C$, and let $C'$ be a face that is topologically contained in $C$ such that Requirement~\ref{req:cycle} is violated with respect to $L$ and $C'$.
Then there are extreme representatives $z_{e1}, z_{e2}$ of $L$ and $C'$ such that $C$ is also implied by $z_{e1}, z_{e2}, x, y$.
\end{lemma}
\begin{proof}
Let $v, t, s, u$ be the defining representatives of $C$ (see Definition~\ref{def:implied} and Figure~\ref{fig:spaths}). Let $z_{e1}$, $z_{e2}$ be any extreme representatives that have representatives $v$ and $t$, respectively, on $C$. By Proposition~\ref{prp:outerplanar}, there can not be $C$-disjoint paths from $z_{e1}$ to both $x$ and $y$. So, without loss of generality, there is no $C$-disjoint path $z_{e1} \leadsto x$. Since $z_{e1}$ has representative $v$ on $C$, a shortest path $z_{e1} \leadsto x$ contains $v$ by Proposition~\ref{prp:cycle:shortest}, and $d(z_{e1}, x) = d(z_{e1}, v) + d(v, x)$. Note that there is a path $z_{e1} \leadsto y$ that contains $v$. If this is not a shortest path, then $d(z_{e1}, y) < d(z_{e1}, v) + d(v, y) = d(z_{e1},v) + d(v,x) = d(z_{e1},x)$, which implies that $z_{e1}$ resolves the pair, a contradiction. Hence, there exists a shortest path $z_{e1} \leadsto y$ that contains $v$. By similar reasoning, there exist shortest paths from $z_{e2}$ to $x$ and to $y$ that contain $t$. Since $z_{e1}$ and $z_{e2}$ do not resolve $x$ and $y$ by assumption, $z_{e1},z_{e2},x,y$ also imply $C$.
\qed\end{proof}

We can now define a configuration for a face $v'$ that has two representatives, a so-called \emph{configuration of type II}.
It specifies a set of extended representatives, and possibly a set of extreme representatives. To be precise, there are two subtypes:

\begin{definition}
Let $v' \in V(T)$ correspond to a face.
A \emph{configuration of type IIa} on $v'$ specifies a set $R \subseteq s(v') \times N_{T}[v']$ that are valid extended representatives for some set $L \subseteq V(G)$ and $|\{\hz : (\hz,w') \in R\}| = 2$, and a set $X \subseteq V(G)$ such that the extended representative of each vertex of $X$ on $v'$ is in $R$. 
A set $L \subseteq V(G)$ \emph{adheres} to a configuration of type IIa if 
there is a set $L' \subseteq L$ of size two such that $L'$ has two representatives on a dual edge $e'=\edge{v'}{w'}$ for some neighbor $w'$ of $v'$ in $T$, and shortest paths from $L'$ to $e'$ intersect a face $y'$ on which $L$ has at least three representatives (cf.\ the case $t_{el} = 0$ on $e'$); moreover, $R$ is equal to the set of extended representatives of $L$ on $v'$ and $X$ is equal to the set of extreme representatives of $L$ and $v'$ that are not in $B(w', v')$.
\end{definition}
The second subtype is complementary to the first subtype.

\begin{definition}
Let $v' \in V(T)$ correspond to a face.
A \emph{configuration of type IIb} on $v'$ specifies a set $R \subseteq s(v') \times N_{T}[v']$ that are valid extended representatives for some set $L \subseteq V(G)$ and $|\{\hz : (\hz,w') \in R\}| = 2$, and a set $X \subseteq V(G)$ such that the extended representative of each vertex of $X$ on $v'$ is in $R$. 
A set $L \subseteq V(G)$ \emph{adheres} to a configuration of type IIb if 
there is \emph{no} set $L' \subseteq L$ of size two such that $L'$ has two representatives on a dual edge $e'=\edge{v'}{w'}$ for some neighbor $w'$ of $v'$ in $T$, and shortest paths from $L'$ to $e'$ intersect a face $y'$ on which $L$ has at least three representatives; moreover, $R$ is equal to the set of extended representatives of $L$ on $v'$ and $X$ is equal to the set of extreme representatives of $L$ and $v'$.
\end{definition}
Both configurations of type II also specify their own subtype.

Recall that we are considering the case that there are two representatives on $v'$, and thus both configurations of type II specify exactly two representatives and thus at most eight extended representatives (\ie $|R| \leq 8$). Also, by the definition of extreme representatives, it is clear that $|X| \leq 4$.

The following lemma shows that with a configuration of type IIa or IIb, we can verify Requirement~\ref{req:cycle} in polynomial time.

\begin{lemma} \label{lem:confII:algo}
Let $v' \in V(T)$ correspond to a face, let $\mc{C}$ be a configuration of type IIa or IIb on $v'$, and let $L \subseteq V(G)$ adhere to $\mc{C}$.
Using only information of $\mc{C}$ we can determine in polynomial time whether $L$ and $v'$ violate Requirement~\ref{req:cycle}.
\end{lemma}
\begin{proof}
Suppose that the configuration has type IIa, and let $w'$ be the promised neighbor in the definition. Let ${\bf t}$ be a boundary condition on $e'=\edge{v'}{w'}$ such that $L$ adheres to ${\bf t}$. Then $t_{el} = 0$. It now follows from Proposition~\ref{prp:tel:three} that Requirement~\ref{req:cycle} is not violated. 

Suppose that the configuration has type IIb. If there is an implied cycle $C$ that violates Requirement~\ref{req:cycle}, then by Lemma~\ref{lem:identify}, $C$ is implied by two extreme representatives $z_{e1}, z_{e2}$ that have different representatives on $v'$, and by some $x, y \in V(G)$. Therefore, we can iterate over all pairs $x, y$ and check for all pairs of extreme landmarks $z_{e1}, z_{e2}$ that have different representatives on $v'$ whether $z_{e1}, z_{e2}, x, y$ imply a cycle that topologically contains $v'$. If they do, then we check whether some extreme representative in $\mc{C}$ resolves $x$ and $y$. Since bifurcation points and shortest paths can be computed in polynomial time, implied cycles can be found in polynomial time. Hence, the algorithm runs in polynomial time.
\qed\end{proof}

The following lemma is an analogue of Lemma~\ref{lem:conf:I} for configurations of type IIa and IIb.

\begin{lemma} \label{lem:conf:II}
Let $v' \in V(T)$ correspond to a face, let $\mc{C}$ be a configuration of type IIa or IIb on $v'$, and let $L \subseteq V(G)$ adhere to $\mc{C}$.
Let ${\bf t}$ be a boundary condition on $\edge{w'}{v'}$ for some neighbor $w'$ of $v'$ in $T$, such that $L$ adheres to ${\bf t}$. Then $\mc{C}$ determines $t_{rl}, t_{ru}$, $t_{eu}$, $t_{el}$, $t_{b}$, and $g(v, L) \cap s(v')$ for all $v \in s(v')$, and all can be computed in polynomial time. 
Finally, all configurations of type IIa and IIb can be enumerated in polynomial time.
\end{lemma}
\begin{proof}
Let $\hz_{1}$ and $\hz_{2}$ denote the two representatives of $L$ on $s(v')$ as given by the configuration. We first show how $\mc{C}$ determines the boundary condition on $e' = \edge{w'}{v'}$. Suppose that $w'$ is a child of $v'$ (similar arguments hold when $w'$ is the parent of $v'$). Then $t_b$ is the set of all $\hz \in s(\edge{w'}{v'})$ for which $(\hz, v')$ is in the configuration; $t_{rl}$ is union of $t_b$ and the set of all $\hz \in s(\edge{w'}{v'})$ for which $(\hz, w')$ is in the configuration; $t_{ru}$ is the union of $t_b$ and the set of representatives on $s((v',w'))$ of all $\hz$ for which $(\hz, u')$ is in the configuration with $u' \neq w'$.

Suppose that $\mc{C}$ has type IIa, and let $y'$ be as in the definition. If $|t_{rl}| = 2$, then $y'$ must be in $B(w', v')$. Then $t_{el} = 0$, and $t_{eu}$ consists of the extreme representatives in $\mc{C}$. Now consider $|t_{rl}| \leq 1$. Then $t_{el}$ contains the extreme landmarks of $\mc{C}$ that are in $B(w', v')$. Also, if $|t_{ru}| = 2$, then $t_{eu} = 0$. Else,
 $|t_{ru}| = 1$ and $t_{eu} = \{\hz_1\} \mbox{ or } \{\hz_2\}$ depending on which one is further from the vertex in $t_{ru}$.

Now suppose that $\mc{C}$ has type IIb.
Then $t_{el}$ is the set of the extreme representatives specified by $\mc{C}$ that are in $B(w', v')$.
If $|t_{ru}| = 2$, then $t_{eu}$ is the set of extreme representatives specified in the configuration, except those in $B(w', v')$.
If $|t_{ru}| = 1$, then there are two subcases. If both $\hz_1, \hz_2$ have an extension $(\hz_i, u')$ in $\mc{C}$ with $u' \neq w'$, then $t_{eu} = \{\hz_1\} \mbox{ or } \{\hz_2\}$ depending on which one is further from the vertex in $t_{ru}$. Otherwise, $t_{eu}$ is the set of extreme representatives of $\mc{C}$ that are not in $B(w', v')$.

We observe that $g(v, L) \cap s(v') = g(v, \{\hz_{1},\hz_{2}\}) \cap s(v')$ for all $v \in s(v')$. Moreover, we note that all the above computations take polynomial time.

As any face has at most four extreme representatives, the total number of configurations is polynomial. Hence, they can be enumerated in polynomial time.
\qed\end{proof}

\subsubsection{Configurations of type III}\label{subp:III}
Suppose that more than two representatives on the face $v'$ are necessary. Corollary~\ref{cor:implied:three:face} implies that any implied cycle containing the face $v'$ satisfies Requirement~\ref{req:cycle}. It would thus seem that a configuration for this case could just consist of three representatives. However, in order to have properties along the lines of Lemma~\ref{lem:conf:I} and~\ref{lem:conf:II}, we need more information.

\begin{definition}\label{def:confIII}
Let $v' \in V(T)$ correspond to a face.
A \emph{configuration of type III} on $v'$ is a set $\mc{C} \subseteq s(v') \times N_{T}[v']$ that are valid extended representatives for some set $L \subseteq V(G)$ such that the associated set $Q = \{\hz : (\hz, w') \in \mc{C} \mbox{ for some $w'$}\}$ satisfies $3 \leq |Q| \leq 6$. We say that a set $L \subseteq V(G)$ \emph{adheres} to a configuration $\mc{C}$ of type~III if 
\begin{enumerate}
\renewcommand{\theenumi}{\roman{enumi}}%
\renewcommand{\labelenumi}{(\roman{enumi})}
\item\label{def:confIII:0} $\mc{C}$ is a subset of the set of extended representatives of $L$ on $v'$.
\item\label{def:confIII:1} $g(v, L) \cap s(v') = g(v, Q) \cap s(v')$.
\item\label{def:confIII:2} For any $z \in L$, the representative of $z$ on $v'$ is on a shortest path between some $\hz_1, \hz_2 \in Q$.
\item\label{def:confIII:3} Let $w'$ be a child of $v'$ and ${\bf t}$ be a boundary condition on $\edge{w'}{v'}$ such that $L$ adheres to ${\bf t}$. Let $R$ be the set $\{\hz: (\hz, u') \in \mc{C} \mbox{ for some $u' \neq w'$}\}$ on $\edge{w'}{v'}$. Then $t_{ru}$ is equivalent\footnote{As explained after Lemma~\ref{lem:midpoint}, we consider $\{v_{1},v_{2}\}$ equal to $\{v_{1},v_{2},v_{e}\}$.} to $R$.
If $w'$ is the parent of $v'$, this condition holds for $t_{rl}$ instead of $t_{ru}$.
\end{enumerate}
\end{definition}

So essentially a configuration of type III determines the values of $g(v, L) \cap s(v')$ and $t_{ru}$. 

We start  by showing that the number of configurations of type~III is polynomial.
\begin{lemma} \label{lem:conf:IIIa}
Let $v' \in V(T)$ correspond to a face.
The number of configurations of type III on $v'$ is $O(|V(G)|^{6})$, and they can be enumerated in polynomial time.
\end{lemma}
\begin{proof}
By definition, $3 \leq |Q| \leq 6$. Using the observation after Definition~\ref{def:extended}, any vertex of $Q$ has a choice of at most $4$ extensions to an extended representative, leading to $2^{4}$ different possibilities. The bound on the number of configurations follows. Enumerating them in polynomial time is straightforward.
\qed\end{proof}

Given a set $L$ that has at least three representatives on a face $v'$, it is not immediately clear whether there exists a configuration that adheres to it.   We shall show that $L$ adheres to the configuration $\mc{C} = \mathsf{ConfIII}(v', L)$  computed according to Algorithm~\ref{alg:conf}.  In order to prove this result, we first analyze some properties of the set $\mc{C}$ computed by Algorithm~\ref{alg:conf}. The intuition behind the algorithm is that we try to find a set of representatives of $L$ on $v'$ that are spread well on $s(v')$.
In the following, we denote the representatives in $Q$ by $\hz_{1}$, $\hz_{2}$, $\hz_{3}$, and (possibly) $\hz_{4}$,  $\hz_{5}$, and $\hz_{6}$, following the notation in Algorithm~\ref{alg:conf}.

\begin{lemma}\label{Claim1}
Let $v' \in V(T)$ correspond to a face, let $L \subseteq V(G)$ have at least three representatives on $v'$, let $\hat{L}$ be the set of representatives of $L$ on $v'$, and let $Q$  be the set computed by Algorithm~\ref{alg:conf} on input $v'$, $L$. Then any $\hz \in \hat{L}$ is on 
a shortest path $\hz_i \leadsto \hz_j$ for some $i, j \in \{1, 2, 3\}$.
\end{lemma}
\begin{proof}
If every representative is on a shortest path $\hz_1 \leadsto \hz_2$, then the claim holds. Otherwise, $\hz_3$ is selected so that it is not on a shortest path $\hz_1 \leadsto \hz_2$. Then the shortest paths $\hz_1 \leadsto \hz_2$ and $\hz_1 \leadsto \hz_3$ intersect only in $\hz_1$, because otherwise $d(\hz_1, \hz_3) > d(\hz_1, \hz_2)$, contradicting the choice of $\hz_1$ and $\hz_2$. Similarly, the shortest paths $\hz_2 \leadsto \hz_1$, $\hz_2 \leadsto \hz_3$ intersect only in $\hz_2$. Hence, the shortest paths cover all vertices in $s(v')$. 
\qed\end{proof}

\begin{algorithm}[t]
\caption{ConfIII --- Compute a configuration of type III}\label{alg:conf}
\begin{algorithmic}[1]
\REQUIRE Face $v'$, set $L \subseteq V(G)$
\STATE Let $\hat{L}$ consist of the representatives of $L$ on $v'$
\STATE Choose $\hz_1, \hz_2 \in \hat{L}$ at maximal mutual distance.
\STATE Choose $\hz_3 \in \hat{L} \setminus \{\hz_1, \hz_2\}$ which is,
  if possible, not on a shortest path $\hz_1 \leadsto \hz_2$, and under that condition if possible not within distance $1/2$ of $\hz_{1}$ or $\hz_{2}$
\STATE $Q \leftarrow \{\hz_1, \hz_2, \hz_3\}$
\IF{the shortest path $\hz_1 \leadsto \hz_2$ is not unique and both paths
contain a vertex of $\hat{L} \setminus \{\hz_1, \hz_2\}$}
\STATE Choose $\hz_4 \in \hat{L} \setminus \{\hz_1, \hz_2\}$ which is on a different shortest path than $\hz_3$, and under that condition if possible not within distance $1/2$ of $\hz_{1}$ or $\hz_{2}$
\STATE $Q \leftarrow Q \cup \{\hz_4\}$
\ENDIF
\FORALL{$i, j \in \{1, 2, 3, 4\}$}
\IF{$\dist(\hz_i, \hz_j) = (|s(v')|-1)/2$ and on the shortest path $\hz_i \leadsto \hz_j$ there is a $\hz_n \in \hat{L}$ but no vertex of $Q$}
\STATE $Q \gets Q \cup \{\hz_n\}$ \label{li:conf-add}
\ENDIF
\ENDFOR
\STATE Let $\mc{C}$ consist of all extended representatives of $L$ on $v'$ which correspond to elements of $Q$
\RETURN $\mc{C}$
\end{algorithmic}
\end{algorithm}

Let $w'$ be a child of $v'$ --- the case where $w'$ is the parent is similar, and not discussed separately --- and let $e' = \edge{w'}{v'}$. Recall that $t_{ru}$ must be the set of representatives of $L \cap V(B(v', w'))$ on $s(e')$ and that $R$ is the set of representatives of $\{\hz: (\hz, u') \in \mc{C} \mbox{ for some $u' \neq w'$}\}$ on $\edge{w'}{v'}$.  Let $z \in L \cap B(v', w')$ have representatives $\hz$ and $v$ on $v'$ and $e'$, respectively.  Since we have the equivalence relation on $t_{ru}$ motivated by Lemma~\ref{lem:midpoint}, we can prove  the following.

\begin{lemma}\label{lem:confIII}
\label{Claim2} 
Let $v' \in V(T)$ correspond to a face, let $L \subseteq V(G)$ have at least three representatives on $v'$ and let $\mc{C}$ be the set computed by Algorithm~\ref{alg:conf} on input $v'$, $L$.
Let $w'$ be a child of $v'$ and let $e' = \edge{w'}{v'}$.
Suppose some $z \in L \cap V(B(v', w'))$ has representative $\hz$ on $v'$ and representative $v$ on $e'$.
Then there is a $(\hz_{n}, u') \in \mc{C}$ such that $u' \neq w'$ and $\hz_{n}$ has representative $v$ on $e'$ or, if $e'$ corresponds to an inner edge $e = \edge{v_{1}}{v_{2}}$ with midpoint $v_{e}$ and $v = v_{e}$, then there are $(\hz_{n}, u'), (\hz_{n'}, u'') \in \mc{C}$ such that $u',u''\not= w'$ and $\hz_{n}$ and $\hz_{n'}$ together have representatives $\{v_{1},v_{2}\}$ on $s(e')$.
\end{lemma}
\begin{proof}
We prove this result by case analysis. 

\medskip
\noindent
\emph{Case 1:}  There is a $\hz_{n} \in Q$ with representative $v$ on $s(e')$, and $\hz_{n} \not= v$ or $\hz = v$.\\
\noindent
Observe that  if $\hz_{n} \not= v$, then $\hz_{n} \not\in s(e')$ --- or $\hz_{n}$ would not have representative $v$ on $s(e')$ --- and the result immediately follows. If $\hz = v$, then $\hz_{n} \not= v$ or $\hz_{n} = v$. The first case was just considered. In the second case, we recall that $z \not\in V(B^{-}(w',v'))$, and the lemma follows.

\medskip
\noindent
\emph{Case 2:}  $e'$ corresponds to an inner edge $e = \edge{v_{1}}{v_{2}}$ and $v$ is equal to the midpoint $v_{e}$ of $e$.\\
\noindent
The only two vertices of $s(v')$ that have representative $v_{e}$ on $s(e')$ are $v_{e}$ and the vertex or midpoint antipodal to $v_{e}$. Note that $\hz \not= v_{e}$, as $z \not\in V(B^{-}(w',v'))$. Hence, $\hz$ is antipodal to $v_{e} = v$. If $\hz \in Q$, then the result is immediate from Case~1. So assume that $\hz \not\in Q$.
By Lemma~\ref{Claim1}, $\hz$ is on a shortest path $\hz_i \leadsto \hz_j$ for $i,j \in \{1,2,3\}$. If $\{\hz_{i}, \hz_{j}\} \cap \{v_{1},v_{2}\} = \emptyset$, then the claimed result is immediate. Otherwise, w.l.o.g.\ $\hz_i = v_1$. Then $\hz_j$ is antipodal to $\hz_i$ and $\hz_i, \hz_j$ are at maximal distance, i.e.~they are $\hz_1$ and $\hz_2$. The path from $\hz_{1}$ to $\hz_{2}$ via $\hz$ contains at least one representative, namely $\hz$, so Algorithm~\ref{alg:conf} includes a representative $\hz_{k}$ on that path in $Q$. Using $\hz_{j}$ and $\hz_{k}$, the lemma follows.

\medskip
\noindent
\emph{Case 3:} $e'$ corresponds to an inner edge $e = \edge{v_{1}}{v_{2}}$ and $v$ is not equal to the midpoint of $e$ (\ie $v \in V(G)$). \\
\noindent
Without loss of generality, $v = v_{1}$. Let $a'$ be the vertex or midpoint antipodal to $v_{1}$ and let $a$ be the vertex or midpoint antipodal to $v_{e}$. Order the vertices and midpoints of $s(v')$ such that $v_{2} \prec v_{e} \prec v_{1} \prec \cdots$. Let $\hz_{n}$ be the vertex of $Q$ that appears first in this ordering and comes after $v_{1}$. 
If $\hz_{n}$ comes after $a'$, then $\hz$ is not on a shortest path $\hz_{i} \leadsto \hz_{j}$ for any $i,j \in \{1,2,3\}$, contradicting Lemma~\ref{Claim1}.
For the same reason, if $\hz_{n} = a'$, then $\{\hz_{1},\hz_{2}\} = \{a',v_{1}\}$. The existence of $\hz$ implies that one of $\hz_{3}, \hz_{4}$ lies on the shortest path $a' \leadsto v_{1}$ that contains $a$, contradicting the choice of $\hz_{n}$.
If $\hz_{n} = a$, then consider whether $v_{1} \in Q$. 
If $v_{1} \in Q$, then either $v_{1} = \hz$ and the result holds by Case~1 or, as $d(v_{1},a) = (|s(v')|-1)/2$ and $\hz$ is on the shortest path $a \leadsto v_{1}$, line~\ref{li:conf-add} of $\mathsf{ConfIII}$ implies that $Q$ must contain a vertex or midpoint on this path, contradicting the choice of $\hz_{n}$. 
If $v_{1} \not\in Q$, then $v_{e} \in Q$, or $\hz$ would not be on a shortest path $\hz_{i} \leadsto \hz_{j}$ for any $i,j \in \{1,2,3\}$, contradicting Lemma~\ref{Claim1}. Then the choice of $\hz_{n}$ implies that $\{a,v_{e}\} = \{\hz_{1},\hz_{2}\}$. Since $\hz$ is on a shortest path $a \leadsto v_{e}$, one of $\hz_{3},\hz_{4}$ is on the same shortest path. As $v_{1}\not\in Q$, this contradicts the choice of $\hz_{n}$. Therefore, $\hz_{n}\not=v_{1}$ has representative $v_{1}$ on $s(e')$, and the lemma follows.

\medskip
\noindent
It remains to prove that the above three cases are exhaustive. Suppose that $w'$ is not a face, \ie $s(w') = \{v\}$ is a cut vertex. Then any vertex or midpoint of $v'$ has representative $v$ on $e'$. Since $|Q| \geq 3$, there is a $\hz_{n} \in Q$ with representative $v$ on $e'$ such that $\hz_{n} \not= v$. Hence, we get Case~1. We may thus assume that $w'$ is a face and that $e'$ corresponds to an inner edge $e = \edge{v_{1}}{v_{2}}$. Then we get Case~2 or Case~3. Therefore, the cases are exhaustive, and the lemma follows.
\qed\end{proof}

Finally, we can prove our main result about the existence of a type III configuration $\mc{C}$ such that $L$ adheres to $\mc{C}$.

\begin{lemma} \label{lem:conf:IIIb}
Let $v' \in V(T)$ correspond to a face and let $L \subseteq V(G)$ have at least three representatives on $v'$. Then there exists a configuration $\mc{C}$ of type III such that $L$ adheres to $\mc{C}$.
\end{lemma}
\begin{proof}
Let $L$ be as in the lemma statement and let $\hat{L}$ be the set of representatives of $L$ on $v'$. We shall show that $L$ adheres to the configuration $\mc{C} = \mathsf{ConfIII}(v', L)$ (see Algorithm~\ref{alg:conf}). We observe that $\mc{C}$ is a set of extended representatives of $L$ on $v'$.
Let $Q$ be as in Algorithm~\ref{alg:conf}. Since $L$ has at least three representatives on $v'$, $|Q| \geq 3$. Note that we only add a vertex to $Q$ in line~\ref{li:conf-add} if two vertices among $\hz_{1},\hz_{2},\hz_{3},\hz_{4}$ have distance $(|s(v')|-1)/2$ and there is no vertex of $Q$ on the shortest path between them. There are at most two such pairs of vertices among $\hz_{1},\hz_{2},\hz_{3},\hz_{4}$. Therefore, $|Q| \leq 6$, and $\mc{C}$ indeed is a configuration of type III.

We now prove that $\mc{C}$ satisfies properties (i)--(iv) of Definition~\ref{def:confIII}. We denote the representatives in $Q$ by $\hz_{1}$, $\hz_{2}$, $\hz_{3}$, and (possibly) $\hz_{4}$,  $\hz_{5}$, and $\hz_{6}$, following the notation in Algorithm~\ref{alg:conf}. Then $\mc{C}$ satisfies property:
\renewcommand{\theenumi}{\roman{enumi}}%
\begin{enumerate}
\item[(\ref{def:confIII:0})] This follows directly from Algorithm~\ref{alg:conf}.
\item[(\ref{def:confIII:1})] Let $w, v \in s(v')$ be neighbors. We will show that $w \not \in g(v, L) \cap s(v')$ if and only if $w \not \in g(v, Q) \cap s(v')$.
Suppose that $w \not \in g(v, L) \cap s(v')$. Then there is a $z \in L$ with $d(z, w) \leq d(z, v)$. By Lemma~\ref{Claim1}, the representative of $z$ on $v'$ is on a shortest path $\hz_i \leadsto \hz_j$ for some $\hz_i, \hz_j$, where $i,j\in\{1,2,3\}$. Then one of $\hz_{i},\hz_{j}$, say $\hz_{i}$, satisfies $d(\hz_i, w) \leq d(\hz_i, v)$. Hence, $w \not \in g(v, Q) \cap s(v')$.
Conversely, suppose that $w \not \in g(v, Q) \cap s(v')$. Then there is a $\hz \in Q$ with $d(\hz, w) \leq d(\hz, v)$. Replace $\hz$ with any vertex $z \in L$ that has representative $\hz$ on $v'$, and the inequality still holds. Hence, $w \not \in g(v, L) \cap s(v')$. Therefore, property (\ref{def:confIII:1}) holds.
\item[(\ref{def:confIII:2})] This follows directly from Lemma~\ref{Claim1}. 
\item[(\ref{def:confIII:3})] Let $w'$ be a child of $v'$ --- the case where $w'$ is the parent is similar, and not discussed separately --- and let $e' = \edge{w'}{v'}$. Recall that $t_{ru}$ must be the set of representatives of $L \cap V(B(v', w'))$ on $s(e')$ and that $R$ is the set of representatives of $\{\hz: (\hz, u') \in \mc{C} \mbox{ for some $u' \neq w'$}\}$ on $\edge{w'}{v'}$. Since the configuration $\mc{C}$ consists of extended representatives of $L$, it is easy to see that $R \subseteq t_{ru}$. 
Let $z \in L \cap B(v', w')$ have representatives $\hz$ and $v$ on $v'$ and $e'$, respectively. Observe that $v \in t_{ru}$ and we thus need to show that $v \in R$. Since we have the equivalence relation on $t_{ru}$ motivated by Lemma~\ref{lem:midpoint}, Lemma~\ref{Claim2} implies that  $v \in R$. Therefore, property (\ref{def:confIII:3}) holds.
\end{enumerate}
The lemma follows.
\qed\end{proof}

We observe that the properties of a configuration of type III can all be derived from the configuration output by Algorithm~\ref{alg:conf}. Therefore, it would seem that we could change the definition of adhering to a configuration of type III to ``$\mc{C}$ is the result of applying Algorithm~\ref{alg:conf} to $v'$ and $L$''. However, it is problematic to ensure that this holds during the dynamic-programming algorithm given below and still have all the properties that we need. Therefore, we stick with the definition of configurations of type III using the properties that we need from it.

As a final result in this section, we show that a configuration of type III determines $t_{eu}$ or $t_{el}$ of the boundary condition ${\bf t}$ of an edge $\edge{v'}{w'}$, depending on whether $w'$ is a child or the parent of $v'$, respectively.

\begin{proposition}\label{prop:conf:III}
Let $v' \in V(T)$ correspond to a face, let $\mc{C}$ be a configuration of type III on $v'$, and let $L \subseteq V(G)$ adhere to $\mc{C}$. If $w'$ is a child of $v'$, then $\mc{C}$ determines $t_{eu}$ of any boundary condition ${\bf t}$ on $e' = \edge{v'}{w'}$ such that $L$ adheres to ${\bf t}$.
If $w'$ is the parent of $v'$, then $\mc{C}$ determines $t_{el}$ of any boundary condition ${\bf t}$ on $e' = \edge{v'}{w'}$ such that $L$ adheres to ${\bf t}$.
\end{proposition}
\begin{proof}
Suppose that $w'$ is a child of $v'$ (the case that $w'$ is the parent is similar and not discussed separately).
If $w'$ is not a face, then $t_{eu} = \emptyset$. So $w'$ is a face.
We know that $\mc{C}$ determines $t_{ru}$ by definition. If $|t_{ru}| \geq 2$, then by definition $t_{eu} = 0$. If $|t_{ru}| = 0$, then $t_{eu} = \emptyset$.
Otherwise, $|t_{ru}| = 1$. Since $L$ has at least three representatives on $v'$, the element of $t_{ru}$ cannot be a midpoint. Then, by definition, $t_{eu}$ is the set with the representative on $v'$ furthest from the vertex in $t_{ru}$. It remains to determine this representative.
Let $s(e') = \{v_1, v_2\}$. Without loss of generality, $t_{ru} = \{v_1\}$. Let $b$ be the vertex or midpoint antipodal to $v_2$. Since $t_{ru} = \{v_1\}$, all representatives of $L$ on $v'$ are on the path $P$ from $v_2$ to $b$ via $v_1$. Let $\hz_1, \hz_2$ be the representatives of $L$ on $v'$ that are at maximal mutual distance---it follows from property (\ref{def:confIII:2}) of Definition~\ref{def:confIII} that they are determined by $\mc{C}$. In particular, $\hz_1, \hz_2$ are on $P$, and all representatives on $v'$ are on a shortest path from $\hz_1$ to $\hz_2$.
Then $t_{eu} = \{\hz_1\} \mbox{ or } \{\hz_2\}$, whichever is at maximal distance from $v_{1}$.
\qed\end{proof}

\subsection{Algorithm} \label{sec:algorithm:algorithm}
We now bring all the pieces of the previous sections together in the final algorithm. The algorithm uses bottom-up dynamic programming over the generalized dual tree $T$, which is rooted at an arbitrary vertex $v'_{r} \in V(T)$. The algorithm consists of several subroutines, which we describe in turn below. Throughout this section, let $G$ be a connected outerplane graph, let $T$ be the generalized dual tree of $G$ rooted at an arbitrary vertex $v'_{r}$, and let $s$ be the annotation of $T$ as defined before.

The main subroutine of the algorithm computes the table $m[v',{\bf t}]$, where $v' \in V(T)\setminus\{v_{r}'\}$ has parent $p'$ and ${\bf t}$ is a boundary condition on the edge $e' = \edge{v'}{p'}$. The value of $m[v',{\bf t}]$ is a smallest set $L' \subseteq V(B(v',p'))$ such that for any set $L \subseteq V(G)$ that satisfies $L' = L \cap V(B(v',p'))$ and that adheres to ${\bf t}$, the following holds:
\begin{enumerate}
\item Requirement~\ref{req:g} is satisfied for any vertex of $V(B^{-}(v',p'))$;
\item Requirement~\ref{req:cycle} is satisfied for any face of $B(v', p')$;
\end{enumerate}
if such a set $L$ exists. Otherwise, the value of $m[v',{\bf t}]$ is $\mbox{\sc nil}$. For notational convenience, we define $|\mbox{\sc nil}| = \infty$ and $\mbox{\sc nil} \cup A = \mbox{\sc nil}$ for any set $A$.

The values of $m[v',{\bf t}]$ are computed in a recursive manner: the computation of $m[v',{\bf t}]$ uses the values of $m[w',{\bf r}^{w'}]$ for children $w'$ of $v'$. The basic idea to compute $m[v', {\bf t}]$ is to consider all configurations for $v'$. Each such configuration $\mc{C}$ may determine a set $L_{v'} \subseteq s(v')$ of landmarks on $s(v')$. We then use a dynamic-programming algorithm that determines for each child $w'$ of $v'$ a boundary condition ${\bf r}^{w'}$ that is compatible with ${\bf t}$ and the configuration $\mc{C}$, such that $L_{v'} \cup\, (\bigcup_{w'} m[w', {\bf r}^{w'}])$ is the smallest set that adheres to $\mc{C}$ and that satisfies all conditions that also hold for $m[v',{\bf t}]$. Iterating over all configurations then gives $m[v',{\bf t}]$. We make this intuitive description more precise below.

Let $v'\in V(T)\setminus\{v_{r}'\}$ with parent $p'$ and let ${\bf t}$ be a boundary condition on the edge $e' = \edge{v'}{p'}$. Observe that it follows from Lemmas~\ref{lem:conf:I}, \ref{lem:conf:II}, and~\ref{lem:conf:IIIa} that all configurations of $v'$ can be enumerated in polynomial time. Let $\mc{C}$ be a particular configuration of $v'$. We need to use this configuration to verify that Requirement~\ref{req:g} and~\ref{req:cycle} hold. For Requirement~\ref{req:g}, we note that if $v'$ is a face, then there are at least two representatives on $v'$ specified by the configuration, and thus $|g(v,L) \cap s(v')| \leq 1$ and there is no violation of Requirement~\ref{req:g}. If $v'$ is not a face, then the configuration is of type I, and ensures that there is no violation. For Requirement~\ref{req:cycle}, note that if $\mc{C}$ is a configuration of type I (i.e.~$v'$ corresponds to a cut or pendant vertex) or if $\mc{C}$ is a configuration of type III (i.e.~$v'$ is a face with at least three representatives), then Requirement~\ref{req:cycle} is trivially satisfied. If $v'$ is a face and $\mc{C}$ is a configuration of type IIa or IIb, then the algorithm must verify that Requirement~\ref{req:cycle} indeed holds. Lemma~\ref{lem:confII:algo} gives a polynomial-time algorithm to do this.

When the algorithm considers a configuration $\mc{C}$ on a face $v'$ and boundary condition ${\bf t}$ on edge $\edge{w'}{v'}$, they must agree, in the sense that there must exist a set of vertices that adheres to both of them. (We assume that there is a resolving set that adheres to ${\bf t}$ and a resolving set that adheres to $\mc{C}$. Otherwise, it is irrelevant whether $\mc{C}$ and ${\bf t}$ agree or not.) We say that $\mc{C}$ and ${\bf t}$ \emph{agree} when the following hold:
\begin{itemize}
\item The conclusions of Lemma~\ref{lem:conf:I} (if $\mc{C}$ is of type I), \ref{lem:conf:II} (if $\mc{C}$ is of type IIa or IIb), or Definition~\ref{def:confIII} and Proposition~\ref{prop:conf:III} (if $\mc{C}$ is of type III) hold.
\item If the configuration specifies an extended landmark $(\hz, w')$, then the boundary condition ${\bf t}$ must have a $\hz \in t_{rl}$ if $w'$ is a child of $v'$, and $\hz \in t_{ru}$ if $w'$ is the parent. If the configuration specifies an extended landmark $(\hz, v')$ with $\hz \in s(\edge{w'}{v'})$, then $\hz \in t_b$.
\item If $\mc{C}$ has type III and the boundary condition ${\bf t}$ specifies that $B(w', v')$ has a landmark with representative $\hz$ on $v'$, then
$\hz$ and $\mc{C}$ satisfy point (\ref{def:confIII:2}) of Definition~\ref{def:confIII}.
\item If $w'$ is the parent of $v'$ and $t_{v_1} = 0$, then the configuration must specify $g(v_1, L) \cap s(v') = \emptyset$. The same holds mutatis mutandis for $t_{v_{2}}$.
\item If $w'$ is a child of $v'$ and $t_{v_1} = 1$, then the configuration must specify $g(v_1, L) \cap s(v') = \emptyset$. The same holds mutatis mutandis for $t_{v_{2}}$.
\end{itemize}

We now determine the boundary conditions ${\bf r}^{w'}$ for all children $w'$ of $v'$. Some components are specified by the configuration (see Lemmas~\ref{lem:conf:I}, \ref{lem:conf:II}, and Definition~\ref{def:confIII} and Proposition~\ref{prop:conf:III}). For some others any value is valid, so we choose one that minimizes $L'$. However, the components $t_{v_1}, t_{v_2}$ for different edges have to be compatible in order to satisfy Requirement~\ref{req:g}, which makes optimization more complicated.

Suppose that children $w'$ and $w''$ share a vertex $v$, \ie $v \in V(B(w',v')) \cap V(B(w'',v'))$. Then, by Requirement~\ref{req:g}, $|g(v, L)| = r^{w'}_{v} + r^{w''}_{v} + |g(v,L) \cap s(v')| \leq 1$ has to hold. We use an index of the dynamic-programming table to ensure that this holds.

We now give the dynamic-programming algorithm. Let the vertices of $s(v')$ be $u_{1},\ldots,u_{\ell}$, so that they appear in this order on the cycle $s(v')$ and, if $v'$ is a face, then $u_{1} \in s(p')$ and $u_{2} \not\in s(p')$. Note that $s(\edge{v'}{w'})$ for any child $w'$ of $v'$ consists of at most two consecutive vertices. Then this also induces an ordering $\prec$ on the children of $v'$, namely $w' \prec x'$ if for some $i < j$, $s(w')$ contains vertex $u_{i}$ and $s(x')$ contains vertex $u_{j}$. If $v'$ is a not a face, then $\prec$ can be chosen arbitrarily. Using $\prec$, we can order the children of $v'$ as $w_{1}',\ldots,w_{k}'$. For each child $w_{i}'$ of $v'$, we will use $v_{1}^{i}$ to denote the vertex $u_{b}$ of $s(v')$, where $b$ is the highest index such that $u_{b} \in s(w_{i}')$. If $|s(w_{i}') \cap s(v')| = 2$, then we use $v_{2}^{i}$ to denote the other vertex of $s(w_{i}') \cap s(v')$.

Consider the children of $v'$ according to the order given above.
Let $n[i, b] = |\bigcup_{j=1}^{i} m[w_j', {\bf r}^{w_j'}]|$ ($0 \leq i \leq l$, $b \in \{0, 1\}$), where the ${\bf r}^{w_j'}$ are chosen so they minimize $n[i, b]$ among all choices for which Requirement~\ref{req:g} is not violated. Furthermore, for any ${\bf r}^{w_j'}$ that contains $r^{w_j'}_{v_1^i}$ we have $r^{w_j'}_{v_1^i} \leq b$.
These values are computed by a simple recursion. If $v_1^i = v_1^{i-1}$, then
\[
n[i, b] = \min_{{\bf r}^{w_i'}, b_0} |m[w_i', {\bf r}^{w_i'}]| + n[i-1, b_0]
\]
where the possible values of ${\bf r}^{w_i'}$ agree with $\mc{C}$ and $b_0 + r^{w_i'}_{v_1^{i}} \leq b$.
If $v_1^i \neq v_1^{i-1}$, then
\[
n[i, b] = \min_{{\bf r}^{w_i'}, b_0} |m[w_i', {\bf r}^{w_i'}]| + n[i-1, b_0]
\]
where possible values of ${\bf r}^{w_i'}$ agree with $\mc{C}$, and $r^{w_i'}_{v_1^{i-1}} + b_0 \leq 1$ (if the boundary condition contains element $r^{w_i'}_{v_1^{i-1}}$).
Finally, $n[\ell, b]$, where $b$ is determined by the boundary condition between $v'$ and its parent, equals the size of the smallest set $L_{v'} \cup\, (\bigcup_{w'} m[w', {\bf r}^{w'}])$ which agrees with the given configuration.
From the same computation we can also obtain the actual set, not only its cardinality.

We can apply the same algorithm, with minor modifications, to the root $v'_{r}$. This leads to our main result.

\begin{theorem} \label{thm:final}
The {\metricdp} problem on outerplanar graphs can be solved in polynomial time.
\end{theorem}
\begin{proof}
The proof proceeds by induction. Assume that $m[w', {\bf r}^{w'}]$ has been correctly computed for all children $w'$ of $v'$ and all valid boundary conditions ${\bf r}^{w}$. Then, given a configuration $\mc{C}$ and boundary condition ${\bf t}$, we can compute (as described above) a smallest set $L_{v'} \cup \bigcup_{w'} m[w', {\bf r}^{w'}]$ that adheres to $\mc{C}$ and ${\bf t}$ and satisfies the relevant conditions. By iterating over all configurations $\mc{C}$, we get $m[v', {\bf t}]$ in polynomial time. The set that is returned when the (modified) algorithm is applied to the root $v_{r}'$ is a smallest set that satisfies Requirement~\ref{req:g} and Requirement~\ref{req:cycle}. By Theorem~\ref{thm:main}, this is a minimum resolving set.
\qed\end{proof}

We give a rough estimate of the running time of the algorithm of Theorem~\ref{thm:final}. 
If the outerplane graph $G$ has $n$ vertices, then the generalized dual tree has $O(n)$ vertices. Each face has $O(n^4)$ configurations of type II. For each of them, verifying Requirement~\ref{req:cycle} following the approach of Lemma~\ref{lem:confII:algo} can be done in $O(n^2)$ time with appropriate pre-calculated tables. Each face also has $O(n^6)$ configurations of type III. With such configurations, verifying Requirement~\ref{req:cycle} is not necessary. However, for each such configuration, the algorithm enumerates $O(n)$ boundary conditions independently, and there are $O(n^4)$ choices for each of them per Lemma~\ref{lem:boundary}.
So processing a face requires $\max(O(n^4)\cdot O(n^2), O(n^6)\cdot O(n) \cdot O(n^4)) = O(n^{11})$ operations, and processing the entire generalized dual tree has time complexity $O(n^{12})$. A more involved analysis could lower this rough bound.

\section{Conclusions and Open Problems} \label{conc}
We have shown  that  {\metricdp} is NP-hard for planar graphs, even when the graph has maximum degree~$6$ (an open problem from 1976). We also gave a polynomial-time algorithm to solve the problem on outerplanar graphs. Our algorithm is based on innovative use of dynamic programming which allows us to deal with the non-bidimensional, global problem of \metricdp.

We  pose some open problems about {\metricdp}. First, it would be nice to extend our results to $k$-outerplanar graphs\footnote{Recall a graph is \emph{$k$-outerplanar} if the graph has a planar embedding such that one can obtain the empty graph by performing the following operation $k$ times: remove all vertices bordering the outer face. Note that this implies that outerplanar graphs are $1$-outerplanar.}. The main obstacle to extending the result is that the separators to be associated with nodes of the computation tree should include faces and edges between consecutive levels. For such separators we lose the crucial property that shortest paths between nodes in different parts cross the separator only once.

Even if the problem turns out to be solvable on $k$-outerplanar graphs by a polynomial-time algorithm, it is not clear that such an algorithm could be used to derive a polynomial-time approximation scheme for {\pmetricdp}. The quest for such an approximation scheme or even for a constant-factor approximation algorithm is an interesting challenge in its own right.

We briefly mention two graph classes that are related to $k$-outerplanar graphs on which the complexity of \metricdp{} is open. First, a problem that could be helpful on the way to understand the complexity of \metricdp{} on $k$-outerplanar graphs is \metricdp{} on irregular grids with or without holes.
Although \metricdp{} on bipartite graphs is $NP$-complete~\cite{EpsteinLW2012}, it seems to be open on grids.
Second, a common generalization of $k$-outerplanar graphs are graphs of bounded treewidth. What is the complexity of \metricdp{} on such graphs?
In particular, it would be very interesting to find out whether \metricdp{} can be formulated as an MSOL-formula.

Generalizing in a different direction, one could consider the weighted version of \metricdp{}. In the paper by Epstein~\etal\cite{EpsteinLW2012}, all the graph classes for which the unweighted version can be solved in polynomial time are also classes for which the weighted version can be solved in polynomial time. Therefore, it is interesting whether the metric dimension of outerplanar graphs can be computed in polynomial time when the given graph is weighted. 

Another interesting line of research is the {\em parameterized complexity} of {\metricdp}. Daniel Lokshtanov~\cite{Lokshtanov09} posed this problem at a Dagstuhl seminar on parametrized complexity. Moreover, he conjectured that the problem could be W[1]-complete. As already mentioned, Hartung and Nichterlein~\cite{HartungN} recently showed that the problem is actually W[2]-complete for the standard parameter (the size of the resolving set) on graphs of maximum degree three, closing this problem. In contrast, Foucaud~\etal\cite{FoucaudMNPV2015,FoucaudMNPV2015b} showed that the problems is fixed-parameter tractable for the standard parameter on interval graphs, and Belmonte~\etal\cite{BelmonteFGR2016} generalized this to all graphs of bounded treelength. However, the parameterized complexity on planar graphs remains open. We hope that the insights of this paper can help to obtain results in this direction. 

\medskip
\noindent
{\bf Acknowledgment:} The authors thank the anonymous reviewers for their helpful suggestions.

\medskip
\noindent
{\em In memoriam of David S.\ Johnson, who suggested the authors to further explore the complexity of {\metricdp}.}

\end{document}